# Distributed Cubature Kalman Filter based on MEEF with Adaptive Cauchy Kernel for State Estimation

Duc Viet Nguyen, Haiquan Zhao, *Senior Member, IEEE*, Jinhui Hu

***Abstract*— Nowadays, with the development of multi-sensor networks, the distributed cubature Kalman filter is one of the well-known existing schemes for state estimation, for which the influence of the non-Gaussian noise, abnormal data, and communication burden are urgent challenges. In this paper, a distributed cubature Kalman filter based on adaptive minimum error entropy with fiducial points (AMEEF) criterion (AMEEF-DCKF) is proposed to overcome the above limitations. Specifically, firstly, in order to solve the influence of various types of non-Gaussian noise and abnormal data, the AMEEF optimization criterion is designed, in which the kernels used are Cauchy kernels with adaptive bandwidth. At the same time, the designed optimization criterion has enhanced the numerical stability and optimized the kernel bandwidth value. Next, in order to address the communication burden problem in multi-sensor networks, where a leader and a follower are distinguished, a distributed algorithm is constructed to achieve an average consensus among these sensors, called leader-follower average consensus (LFAC). Additionally, the convergence proof of the average consensus algorithm and the computational complexity analysis of the AMEEF-DCKF algorithm are also presented. Finally, through a 10-node sensor network, the effectiveness of the proposed algorithm is demonstrated in estimating the state of the power system and navigating land vehicles in complex environments.**

## I. Introduction

With the ability to aggregate information powerfully in space and time, multi-sensor networks have been widely used in practical applications such as cooperative localization, Internet of Things, navigation, state estimation, etc [1-3]. Recently, it can be observed that the trend has shifted from traditional centralized architectures to distributed ones due to the latter's superior robustness and scalability [4]. With distributed architecture, the communication burden has been eliminated, which makes it widely used in practice [5]. Within distributed multi-sensor networks, a fundamental problem that must be addressed is the consensus problem, which is generally categorized into leaderless and leader-follower types [6-8]. Currently, in leader-follower consensus, consensus with multiple leaders is receiving attention because of the fault tolerance it achieves [9,10]. However, these studies are limited to driving the followers into the convex hull formed by the leaders' states. To move beyond this limitation, this paper aims to design a consensus algorithm that enables the followers to converge to the average value of the leaders' states.

In state estimation based on distributed multi-sensor networks, distributed Kalman filters (DKF) have always been a popular technical solution, due to their simplicity but effectiveness [11,12]. Comparing the performance of nonlinear DKF variants, distributed cubature Kalman filters (DCKF) have achieved better results under the same estimation conditions [13]. In addition, traditional Kalman filters built on the Gaussian assumption may not be suitable in applications in which there is non-Gaussian noise, impulse noise, etc. Recently, distributed Kalman filters using information theoretic learning (ITL) such as maximum correntropy criterion (MCC), minimum error entropy (MEE) as optimization criteria, have been introduced [14-17]. With the ability to handle non-Gaussian noise and outliers, these studies have immediately attracted the attention. The MEE-based DKF (MEE-DKF) with good modeling ability of error entropy has shown better estimation results than MCC [4,18]. Nonetheless, the MEE criterion has limitations in terms of numerical stability [19]. While various solutions, such as local approximations [20] and regularization techniques [21], have been proposed to mitigate this issue, the most successful approach has been the fusion of MCC principles into MEE, yielding the MEE with fiducial points (MEEF) criterion. In MEEF, the error probability density function is automatically centered at zero, significantly enhancing robustness. Consequently, the Kalman filters developed on the MEEF criterion exhibit remarkable capability in handling non-Gaussian noise characterized by complex distributions or significant outliers [22,23].

It can be observed that the DKF based on the above information criteria all use Gaussian kernels, which are known to be very sensitive to the bandwidth size and require a lot of experimentation [24]. Furthermore, the use of Gaussian kernels can lead to numerical instability during Cholesky decomposition, where the occurrence of singular matrices may disrupt the estimation process [25]. In order to solve the problem of choosing the optimal size of the kernel bandwidth, optimization algorithms have been applied to automatically calculate and select [26,27]. However, this solution can significantly increase the size and computation time of the estimation algorithms. As a more efficient alternative, recent studies have explored the use of Cauchy kernels, which demonstrate considerably lower sensitivity to bandwidth

This work was partially supported by the National Natural Science Foundation of China (grant: 62171388, 61871461, 61571374) and Major Special Project of China Railway Group Limited (Source -2025- Special Project -001) (*Corresponding author*: Haiquan Zhao).

Haiquan Zhao (e-mail: hqzhao_swjtu@126.com), Duc Viet Nguyen (e-mail: ducvietpkkq@gmail.com), and Jinhui Hu (e-mail: Hujinhui@my.swjtu.edu.cn) are with the Key Laboratory of Magnetic Suspension Technology and Maglev Vehicle, Ministry of Education, School of Electrical Engineering, Southwest Jiaotong University, Chengdu, China.

parameters, to replace the traditional Gaussian kernels [28-30]. Additionally, it should be noted that the estimation process is dynamic, and the noise distribution may change over time. Therefore, an adaptive strategy that dynamically adjusts the kernel bandwidth to match the prevailing noise characteristics can significantly improve estimation accuracy compared to using a fixed bandwidth throughout the process [31,32].

In summary, this work is motivated by the need to address key challenges in distributed nonlinear estimation under non-Gaussian noise, namely: (1) robustness to abnormal data and time-varying noise distributions; (2) numerical stability issues, including singularity problems during matrix decomposition; and (3) sensitivity to kernel bandwidth parameters. To tackle these challenges, we propose a novel distributed adaptive estimation algorithm, termed the adaptive minimum error entropy with fiducial points-based distributed Cubature Kalman filter (AMEEF-DCKF). The main contributions of this paper are as follows:

1) A leader-follower average consensus (LFAC), leveraging the Push-Sum protocol, is designed. This algorithm enables the followers to achieve consensus on the average value of the leaders' states, thereby alleviating the communication burden inherent in multi-sensor networks.

2) The proposed AMEEF criterion employs an adaptive Cauchy kernel to enhance robustness against non-Gaussian noise and abnormal data. The use of this kernel concurrently improves numerical stability by preventing matrix singularity issues and reducing sensitivity to bandwidth selection. This adaptability aligns well with the time-varying characteristics of practical noise distributions.

3) The AMEEF-DCKF algorithm for state estimation of nonlinear systems within a multi-sensor network, integrating the designed LFAC algorithm, is proposed. Through numerical experiments in complex environments with objects such as power systems and land vehicles, the excellent performance of the proposed algorithm is demonstrated.

4) The convergence proof of the LFAC algorithm and the computational complexity of the AMEEF-DCKF algorithm are also given.

In order to clarify the above contributions, this paper is organized as follows: In Section II, we will summarize MEEF and design the LFAC algorithm. The construction sequence of the proposed algorithm AMEEF-DCKF is presented in Section III. Section IV provides a convergence proof of the LFAC algorithm and analyzes the computational complexity of AMEEF-DCKF. In Section V, numerical experiments are given to illustrate the excellent performance of AMEEF-DCKF. In Section VI, the conclusions are given.

## II. MEEF Criterion and Leader-Follower Average Consensus Algorithm

### A. MEEF criterion

In ITL, Renyi's quadratic entropy [4,14,17] with error variable $e$ and probability density function (PDF) $p_e(.)$, is defined as follows:

$$H_2(e) = -\log_2\left(\int \left(p_e(e)\right)^2 de\right) \tag{1}$$

By minimizing Eq.(1), the MEE criterion is inferred:

$$V(e) = \int \left(p_e(e)\right)^2 de \tag{2}$$

In practice, the $p_e(e)$ is unknown, by combining Parzen's estimator and utilizing samples $\{e_i\}_{i=1}^N$ [20], the PDF is estimated as follows:

$$\hat{p}_e(e) = \frac{1}{N}\sum_{i=1}^{N} C_\sigma(e-e_i) \tag{3}$$

where $C_\sigma(e) = 1/\left(1+e^2/\sigma\right)$: Cauchy kernel function with kernel width σ; $N$: the number of errors. Substituting Eq.(3) into Eq.(2), the MEE criterion involves maximizing:

$$\hat{V}(e) = \frac{1}{N}\sum_{i=1}^{N}\hat{p}_e(e_i) = \frac{1}{N^2}\sum_{i=1}^{N}\sum_{j=1}^{N} C_\sigma(e_i-e_j) \tag{4}$$

It can be observed that after optimization, the error peak may not be located at zero, because the MEE criterion only focuses on minimizing the difference. Different from the traditional methods, to solve this problem, an enhanced error form $E[e_0,\mathbf{e}]$, in which $\mathbf{e}=[e_1,\ldots,e_N]$ represents the error samples, $e_0$ acts as a fiducial point, is constructed. With this automated approach, it is equivalent to adding the MCC criterion to the MEE [22,23]. Performing the maximization and ignoring the coefficients, the MEEF criterion is obtained as:

$$J = \eta\sum_{j=1}^{N} C_{\sigma_1}(e_j) + (1-\eta)\sum_{i=1}^{N}\sum_{j=1}^{N} C_{\sigma_2}(e_j-e_i) \tag{5}$$

where $\eta$ ($0\leq \eta\leq 1$) is the mixture coefficient; $\sigma_1$, $\sigma_2$ are the kernel width.

*Remark 1*: When $\eta$=1, the MEEF criterion degenerates to the MCC criterion [15,16]; when $\eta$=0, the MEEF criterion degenerates to the MEE criterion [17,21].

### B. Leader-follower average consensus algorithm

Consider a multi-sensor network ($n$ sensors) consisting of $m$ stationary leaders and ($n$-$m$) followers, where they interact through a network described by a directed graph £, with the dynamics described as follows [9]:

$$\beta_i(t+1) = \alpha_i\beta_i(t) + u_i(t) \tag{6}$$

where $\alpha_i$ is the weight coefficient; $u_i(t)$ is the control input for agent $i$ at the $t^{th}$ iteration.

Suppose the update protocol used by the sensors as:

$$u_i(t) = \begin{cases} \sum_{j\in\aleph_i^{in}} \alpha_{ij}\beta_i(t) & i\in F \\ 0 & i\in L \end{cases} \tag{7}$$

where $\aleph_i^{in}$ is the all in-neighbours; $\beta_L(t)\in\mathbb{R}^m$; $\beta_F(t)\in\mathbb{R}^{n-m}$ $F$ and $L$ denote the index set of followers and leaders, respectively.

Then, the dynamics of the sensors in Eq.(6) can be rewritten in matrix form as follows [10]:

$$\begin{bmatrix} \beta_F(t+1) \\ \beta_L(t+1) \end{bmatrix} = \begin{bmatrix} A_1 & A_2 \\ 0 & I_m \end{bmatrix} \begin{bmatrix} \beta_F(t) \\ \beta_L(t) \end{bmatrix} \quad (8)$$

$$A = \begin{bmatrix} A_1 & A_2 \\ 0 & I_m \end{bmatrix} \quad (9)$$

where $A \in \mathbb{R}^{n\times n}$ ; $A^{ij} = \alpha_{ij}$ with $i \neq j$ and $A^{ii} = \alpha_i$ is the weighted adjacency matrix.

Suppose that $A_1 = I_{n-m} - \alpha L_1$ and $A_2 = -\alpha L_2$ satisfy the protocol in Eq.(8), in which $L_1$ and $L_2$ are the weighted Laplacian matrices associated with the subgraphs $£^F$ and $£^{FL}$, respectively. Then, the followers will converge to the convex hull formed by the stationary leaders if and only if the step size satisfies ( $\lambda_i$ : characteristic value):

$$\alpha < \min_{\lambda_i \in \Lambda(L_1)} \frac{2\,\mathrm{Re}(\lambda_i)}{\mathrm{Re}^2(\lambda_i) + \mathrm{Im}^2(\lambda_i)} \quad (10)$$

and the final state of all the followers is given by $-L_1^{-1} L_2 \beta_L(0)$ [35].

A known limitation of conventional approaches is that follower states converge to a point within the convex hull of the leaders' states. While the leaders may be stationary, the followers fail to reach consensus on a common value. In order to solve this problem, the protocol Push-Sum is applied [36].

Consider that each sensor will contain two pieces of information, including the sum vector $s_j(t-1)$ with $s(0) = x(0)$ and weight vector $\omega_j(t-1)$ with $\omega(0) = \mathbf{I}_n$ received from all in-neighbours $j \in \aleph_i^{in}$, calculated by:

$$s_i(t) = A^{ii} s_i(t-1) + \sum_{j\in\aleph_i^{in}} A^{ij} s_j(t-1) \quad (11)$$

$$\omega_i(t) = A^{i}\omega_i(t-1) + \sum_{j\in\aleph_i^{in}} A^{ij} \omega_j(t-1) \quad (12)$$

Then, the state vector of each sensor is updated by:

$$\beta_i(t) = \frac{s_i(t)}{\omega_i(t)} \quad ;\forall i \in F \quad (13)$$

Each sensor updates its status until it satisfies the stop criteria:

$$E_i(t) = |e_i(t)| \leq \gamma \quad ;\forall i \in F \quad (14)$$

where $e_i(t) = \beta_i(t) - \beta_i(t-1)$ and $\gamma$ is the threshold.

The states of the followers converge to the average state of the leader, which is the goal of this average consensus algorithm. The mathematical proof will be given in the next section.

$$\lim_{t\to\infty} \beta_i(t) \to \frac{1}{m} \sum_{j\in L} \beta_j(t) \quad ;\forall i \in F \quad (15)$$

## III. AMEEF-DCKF Algorithm

### *A. Derivation of MEEF-CKF*

Consider a nonlinear system, for which the state vector $\mathbf{x}(t) \in \mathbb{R}^n$ and the measurement vector $\mathbf{y}(t) \in \mathbb{R}^m$ at discrete time *t* are described as follows:

$$\mathbf{x}(t) = \mathbf{f}(\mathbf{x}(t-1)) + \mathbf{q}(t-1) \quad (16)$$

$$\mathbf{y}(t) = \mathbf{h}(\mathbf{x}(t)) + \mathbf{r}(t) \quad (17)$$

where **h**(.) and **f**(.) are the measurement and state transition functions, respectively; $\mathbf{r}(t) \in \mathbb{R}^m$ and $\mathbf{q}(t-1) \in \mathbb{R}^n$ denote the measurement noise and zero-mean process noise with covariance matrices **R**(*t*) and **Q**(*t-1*), respectively. Determine the state estimate at *t*-1 in terms of $\hat{\mathbf{x}}(t-1)$ and the corresponding error covariance matrix $\mathbf{P}(t-1) = \mathbf{S}(t-1)\mathbf{S}^T(t-1)$, in which **S**(*t*-1) represents the Cholesky decomposition of **P**(*t*-1).

The proposed algorithm performs the state estimation in two steps:

#### *1) Prediction Step*

Calculate the sample points $\boldsymbol{\delta}_i(t-1)$ :

$$\boldsymbol{\delta}_i(t-1) = \mathbf{S}(t-1)\boldsymbol{\Delta}_i + \hat{\mathbf{x}}(t-1) \quad (18)$$

$$\boldsymbol{\Delta}_i = \begin{cases} \sqrt{n}\mathbf{b}_i & \text{for } i = 1,2,..,n \\ -\sqrt{n}\mathbf{b}_i & \text{for } i = n+1, n+2,..,2n \end{cases} \quad (19)$$

where $\mathbf{b}_i$ is the unit vector.

Through the state transition function, the cubature points are obtained as:

$$\xi_i(t|t-1) = \mathbf{f}(\boldsymbol{\delta}_i(t-1)) \quad ;i = 1,2,...,2n \quad (20)$$

The predicted state $\hat{\mathbf{x}}(t|t-1)$ and the covariance matrix $\mathbf{P}_{xx}(t|t-1)$ are calculated by:

$$\hat{\mathbf{x}}(t|t-1) = \frac{1}{2n} \sum_{i=1}^{2n} \xi_i(t|t-1) \quad (21)$$

$$\mathbf{P}_{xx}(t|t-1) = \frac{1}{2n} \sum_{i=1}^{2n} \hat{\xi}_i(t|t-1)\hat{\xi}_i^T(t|t-1) + \mathbf{Q}(t-1) \quad (22)$$

where $\hat{\xi}_i(t|t-1) = \xi_i(t|t-1) - \hat{\mathbf{x}}(t|t-1)$.

#### *2) Time Update*

Calculate sample points $\boldsymbol{\delta}_i(t|t-1)$ by:

$$\boldsymbol{\delta}_i(t|t-1) = \mathbf{S}(t|t-1)\boldsymbol{\Delta}_i + \hat{\mathbf{x}}(t|t-1) \quad (23)$$

Through the measurement function, the cubature points are obtained:

$$\xi_i(t) = \mathbf{h}(\boldsymbol{\delta}_i(t|t-1)) \quad ;i = 1,2,...,2n \quad (24)$$

Then, $\hat{\mathbf{y}}(t|t-1)$ and $\mathbf{P}_{xy}(t|t-1)$ are the predicted measurement vector and cross-covariance matrix, respectively, obtained by:

$$\hat{\mathbf{y}}(t|t-1) = \frac{1}{2n}\sum_{i=1}^{2n}\xi_i(t) \tag{25}$$

$$\mathbf{P}_{xy}(t|t-1) = \frac{1}{2n}\sum_{i=1}^{2n}\hat{\xi}_i(t|t-1)\hat{\xi}_i^T(t) \tag{26}$$

where $\hat{\xi}_i(t) = \xi_i(t) - \hat{\mathbf{y}}(t|t-1)$.

To continue to complete the measurement update, a linear regression model is applied [15,17,22], in which the state prediction error and measurement function are combined:

$$\begin{bmatrix} \hat{\mathbf{x}}(t|t-1) \\ \mathbf{y}(t) \end{bmatrix} = \begin{bmatrix} \mathbf{x}(t) \\ \mathbf{h}(\mathbf{x}(t)) \end{bmatrix} + \begin{bmatrix} -\mathbf{\Gamma}(t|t-1) \\ \mathbf{r}(t) \end{bmatrix} \tag{27}$$

where $\mathbf{\Gamma}(t|t-1) = \mathbf{x}(t) - \hat{\mathbf{x}}(t|t-1)$ is the prediction error.

Based on the references [37], the statistical linearization technique is used to approximate the measurement function as:

$$\mathbf{y}(t) = \hat{\mathbf{y}}(t|t-1) + \mathbf{S}(t)\mathbf{\Gamma}(t|t-1) + \mathbf{r}(t) + \mathbf{v}(t) \tag{28}$$

$$\mathbf{S}(t) = \left(\mathbf{P}_{xx}(t|t-1)\mathbf{P}_{xy}(t|t-1)\right)^T \tag{29}$$

where $\mathbf{v}(t)$ is the statistical linearization error [22].

Combining Eq.(27) and Eq.(28) into a matrix as:

$$\begin{bmatrix} \hat{\mathbf{x}}(t|t-1) \\ \mathbf{y}(t) - \hat{\mathbf{y}}(t|t-1) + \mathbf{S}(t)\hat{\mathbf{x}}(t|t-1) \end{bmatrix} = \begin{bmatrix} \mathbf{I}_n \\ \mathbf{S}(t) \end{bmatrix}\mathbf{x}(t) + \kappa(t) \tag{30}$$

where $\kappa(t) = \left[-\mathbf{\Gamma}^T(t|t-1) \quad \mathbf{r}^T(t) + \mathbf{v}^T(t)\right]^T$, $\mathbf{I}_n$ is the unity matrix.

It is supposed that the covariance matrix $E\left[\kappa(t)\kappa^T(t)\right]$ of the augmented noise is positive definite [15-17,22]. The covariance matrix of in Eq.(30) is calculated by:

$$\begin{aligned} E\left[\kappa(t)\kappa^T(t)\right] &= \mathbf{\Xi}(t)\mathbf{\Xi}^T(t) \\ &= \begin{bmatrix} \mathbf{\Xi}_p(t|t-1)\mathbf{\Xi}_p^T(t|t-1) & 0 \\ 0 & \mathbf{\Xi}_r(t)\mathbf{\Xi}_r^T(t) \end{bmatrix} \end{aligned} \tag{31}$$

where $\mathbf{\Xi}(t)$, $\mathbf{\Xi}_p(t|t-1)$ and $\mathbf{\Xi}_r(t)$ obtained by Cholesky decomposition of $E\left[\kappa(t)\kappa^T(t)\right]$, $\mathbf{P}_{xx}(t|t-1)$ and $E\left[(\mathbf{r}(t)+\mathbf{v}(t))(\mathbf{r}(t)+\mathbf{v}(t))^T\right]$. Besides, a technique for stabilizing the Cholesky decomposition condition was introduced in [42]. Then left multiplying $\mathbf{\Xi}^{-1}(t)$ on both sides of (26) obtains:

$$\mathbf{d}(t) = \mathbf{W}(t)\mathbf{x}(t) + \mathbf{e}(t) \tag{32}$$

$$\begin{aligned} \mathbf{d}(t) &= \mathbf{\Xi}^{-1}(t)\begin{bmatrix} \hat{\mathbf{x}}(t|t-1) \\ \mathbf{y}(t) - \hat{\mathbf{y}}(t|t-1) + \mathbf{S}(t)\hat{\mathbf{x}}(t|t-1) \end{bmatrix} \\ &= \left[d_1(t), d_2(t), ..., d_N(t)\right]^T \end{aligned} \tag{33}$$

$$\mathbf{W}(t) = \mathbf{\Xi}^{-1}(t)\begin{bmatrix} \mathbf{I}_n \\ \mathbf{S}(t) \end{bmatrix} = \left[\mathbf{w}_1(t), \mathbf{w}_2(t), ..., \mathbf{w}_N(t)\right]^T \tag{34}$$

$$\begin{aligned} \mathbf{e}(t) &= \mathbf{\Xi}^{-1}(t)\begin{bmatrix} -\left(\mathbf{x}(t) - \hat{\mathbf{x}}(t|t-1)\right) \\ \mathbf{r}(t) + \mathbf{v}(t) \end{bmatrix} \\ &= \left[e_1(t), e_2(t), ..., e_N(t)\right]^T \end{aligned} \tag{35}$$

Because of $E\left[\mathbf{e}(t)\mathbf{e}^T(t)\right] = \mathbf{I}_N$, the $\mathbf{e}(t)$ residual error is white [15].

According to the MEEF criterion, the cost function is written as follows:

$$J_{MEEF} = \eta\sum_{j=1}^{N} C_{\sigma_1}\left(e_j(t)\right) + (1-\eta)\sum_{i=1}^{N}\sum_{j=1}^{N} C_{\sigma_2}\left[e_j(t) - e_i(t)\right] \tag{36}$$

where $e_i(t) = d_i(t) - \mathbf{w}_i(t)\mathbf{x}(t)$ ; $d_i(t)$ represent the $i^{th}$ element of $\mathbf{e}(t)$ and $\mathbf{d}(t)$, respectively; $\mathbf{w}_i(t)$ represent the $i^{th}$ row of $\mathbf{W}(t)$. The optimal estimate can be obtained by calculating $\hat{\mathbf{x}}(t) = \arg\max_{\mathbf{x}(t)}\left[J(\mathbf{x}(t))\right]$ and $N = n + m$. Taking the derivative of $J(\mathbf{x}(t))$ with respect to $\mathbf{x}(t)$ be zero:

$$\frac{\partial J_{MEEF}}{\partial \mathbf{x}(t)} = \mathbf{W}^T(t)\Theta(t)\mathbf{d}(t) - \mathbf{W}^T(t)\Theta(t)\mathbf{W}(t)\mathbf{x}(t) = 0 \tag{37}$$

where

$$\begin{cases} \mathbf{\Theta}(t) = \eta_1\mathbf{\Omega}(t) + \eta_2\left[\mathbf{\Phi}(t) - \mathbf{\Psi}(t)\right] \\ \left[\mathbf{\Phi}(t)\right]_{ij} = \begin{cases} \left[\mathbf{\Phi}(t)\right]_{ii} = \sum_{j=1}^{N} C_{\sigma_2}\left(e_j(t) - e_i(t)\right) & ; i = j \\ 0 & ; i \neq j \end{cases} \\ \left[\mathbf{\Psi}(t)\right]_{ij} = C_{\sigma_1}\left(e_j(t) - e_i(t)\right) \\ \left[\mathbf{\Omega}(t)\right]_{ij} = \begin{cases} C_{\sigma_1}\left(e_j(t) - e_i(t)\right) & ; i = j \\ 0 & ; i \neq j \end{cases} \\ \eta_1 = \eta / \sigma_1^2 \\ \eta_2 = 1 - \eta / \sigma_2^2 \end{cases} \tag{38}$$

Similar to studies [19,22,23], the solution of Eq.(37) is estimated by a fixed-point iteration algorithm:

$$\mathbf{x}(t) = \left(\mathbf{W}^T(t)\mathbf{\Pi}(t)\mathbf{W}(t)\right)^{-1}\left(\mathbf{W}^T(t)\mathbf{\Pi}(t)\mathbf{d}(t)\right) \tag{39}$$

With $\mathbf{\Pi}(t) = \eta_1\mathbf{\Omega}_t + \eta_1\left[\mathbf{\Phi}^T(t)\mathbf{\Phi}(t) + \mathbf{\Psi}^T(t)\mathbf{\Psi}(t)\right]$ and the matrix $\mathbf{\Pi}(t)$ can also be rewritten as follows:

$$\mathbf{\Pi}(t) = \begin{bmatrix} \mathbf{\Pi}_{xx}(t) & \mathbf{\Pi}_{yx}(t) \\ \mathbf{\Pi}_{xy}(t) & \mathbf{\Pi}_{yy}(t) \end{bmatrix} \tag{40}$$

where
$\mathbf{\Pi}_{xx}(t) \in \mathbb{R}^{n\times n}; \mathbf{\Pi}_{xy}(t) \in \mathbb{R}^{m\times n}; \mathbf{\Pi}_{yx}(t) \in \mathbb{R}^{n\times m}; \mathbf{\Pi}_{yy}(t) \in \mathbb{R}^{m\times m}$.

Observing Eq. (39), it can be seen that it is equivalent to a function of the variable x(t), which can be obtained by using a fixed-point iteration method. Applying the matrix inverse lemma and according to references [22,23], Eq.(39) can be rewritten as ( $\hat{\mathbf{x}}(t|t)$ is the result obtained of $\mathbf{x}(t)$ via fixed point iteration):

$$\hat{\mathbf{x}}(t|t) = \hat{\mathbf{x}}(t|t-1) + \widehat{\mathbf{K}}(t)\left(\mathbf{y}(t) - \hat{\mathbf{y}}(t|t-1)\right) \tag{41}$$

$$\widehat{\mathbf{K}}(t) = \begin{bmatrix} \widehat{\mathbf{P}}(t|t-1) + \mathbf{S}^T(t)\widehat{\mathbf{P}}_{xy}(t|t-1) + \widehat{\mathbf{P}}_{yx}(t|t-1)\mathbf{S}(t) + \\ +\mathbf{S}^T(t)\widehat{\mathbf{R}}_{yy}(t)\mathbf{S}(t) \end{bmatrix}^{-1} \times\left[\widehat{\mathbf{P}}_{yx}(t|t-1) + \mathbf{S}^T(t)\widehat{\mathbf{R}}_{yy}(t)\right] \tag{42}$$

$$\begin{cases} \widehat{\mathbf{P}}(t|t-1) = \left(\mathbf{\Xi}_p^{-1}(t|t-1)\right)^T \mathbf{\Pi}_{xx}(t)\mathbf{\Xi}_p^{-1}(t|t-1) \\ \widehat{\mathbf{P}}_{xy}(t|t-1) = \left(\mathbf{\Xi}_r^{-1}(t)\right)^T \mathbf{\Pi}_{xy}(t)\mathbf{\Xi}_p^{-1}(t|t-1) \\ \widehat{\mathbf{P}}_{yx}(t|t-1) = \left(\mathbf{\Xi}_p^{-1}(t|t-1)\right)^T \mathbf{\Pi}_{yx}(t)\mathbf{\Xi}_r^{-1}(t) \\ \widehat{\mathbf{R}}_{yy}(t|t-1) = \left(\mathbf{\Xi}_r^{-1}(t)\right)^T \mathbf{\Pi}_{yy}(t)\mathbf{\Xi}_r^{-1}(t) \end{cases} \tag{43}$$

where $\widehat{\mathbf{K}}(t)$ is the gain matrix.

Finally, the covariance matrix $\widehat{\mathbf{P}}(t|t)$ can be updated by:

$$\widehat{\mathbf{P}}(t|t) = \left[\mathbf{I} - \widehat{\mathbf{K}}(t)\mathbf{S}(t)\right]\widehat{\mathbf{P}}(t|t-1)\left[\mathbf{I} - \widehat{\mathbf{K}}(t)\mathbf{S}(t)\right]^T + \widehat{\mathbf{K}}(t)\mathbf{R}(t)\widehat{\mathbf{K}}^T(t) \tag{44}$$

*Remark 2*: The estimated performance of algorithms based on the ITL criterion is directly affected by the value of kernel bandwidth [33]. If these values are too large, the performance obtained is only equivalent to that based on the Gaussian assumption. If these values are too small, it will not be possible to handle non-Gaussian noises, and the fixed-point iteration will diverge [34]. Therefore, in order to achieve better dynamic estimation performance, where the influence noise has a dynamic distribution, the values of those coefficients also need to be flexibly variable.

### *B. Cauchy kernel with adaptive bandwidth*

In order to optimize the dynamic estimation process, this study applies an adaptive adjustment principle: employ a wide bandwidth for noise with near-Gaussian characteristics and a narrower one for noise with near-non-Gaussian characteristics. To be consistent with reality, the measurement noise covariance matrix can be described as follows:

$$\tilde{\mathbf{R}}(t) = E\left(\mathbf{r}(t)\mathbf{r}^T(t)\right) \neq \mathbf{R}(t) \tag{45}$$

According to remark 2 and reference [38], the RMSE of the MEEF-based filter will be less than or equal to the MMSE criterion if and only if the condition is satisfied:

$$\tilde{\mathbf{R}}(t) \le C^{-1}(t)\left[\mathbf{P}(t) - \mathbf{R}(t)\right]C(t) + 2\mathbf{R}(t)C^{-1}(t) \tag{46}$$

$$C(t) = diag\left[C_\sigma\left(\left\|\widehat{\mathbf{y}}_1(t)\right\|^2_{R_1^{-1}(t)}\right), \ldots, C_\sigma\left(\left\|\widehat{\mathbf{y}}_m(t)\right\|^2_{R_m^{-1}(t)}\right)\right] \tag{47}$$

where: $\widehat{\mathbf{y}}(t) = \mathbf{y}(t) - \hat{\mathbf{y}}(t|t-1)$ : the innovation term; $\widehat{\mathbf{y}}_i(t)$ : the $i^{th}$ value of the innovation term; $\mathbf{R}_i(t)$: the $i^{th}$ diagonal element of $\mathbf{R}(t)$.

Consider the elements on the $i^{th}$ diagonal in Eq.(46) to obtain:

$$\tilde{\mathbf{R}}_i(t) \le \left[\mathbf{P}_i(t) - \mathbf{R}_i(t)\right] + 2\mathbf{R}(t)C_\sigma\left(\left\|\widehat{\mathbf{y}}_i(t)\right\|^2_{R_i^{-1}(t)}\right)^{-1} \tag{48}$$

where $\tilde{\mathbf{R}}_i(t)$ and $\mathbf{P}_{i,yy}$ are the the $i^{th}$ diagonal element of $\tilde{\mathbf{R}}(t)$ and $\mathbf{P}_{yy}$, respectively.

By substituting the Cauchy kernel formula into Eq.(48) and rearranging, can be obtained:

$$\frac{\tilde{\mathbf{R}}_i(t) - \left[\mathbf{P}_i(t) - \mathbf{R}_i(t)\right]}{2\mathbf{R}_i(t)} \le 1 + \frac{\left\|\widehat{\mathbf{y}}_i(t)\right\|^2_{R_i^{-1}(t)}}{\sigma(t)} \tag{49}$$

This is the condition corresponding to the Cauchy kernel derived from Eq.(46) so that the error of the MEEF-based filter is better than the MMSE criterion. Since $\left[\tilde{\mathbf{R}}_i(t) - \left(\mathbf{P}_i(t) - \mathbf{R}_i(t)\right)\right] / \left[2\mathbf{R}_i(t)\right] \ge 1$, thus Eq.(49) can be rewritten by:

$$\sigma(t) \le \frac{\left\|\widehat{\mathbf{y}}_i(t)\right\|^2_{R_i^{-1}(t)}}{\left[\tilde{\mathbf{R}}_i(t) - \left(\mathbf{P}_i(t) - \mathbf{R}_i(t)\right)\right] / \left[2\mathbf{R}_i(t)\right] - 1} \tag{50}$$

*Remark 3*. It can be concluded that the performance of the MEEF-CKF algorithm is superior to traditional CKF if and only if the Cauchy kernel bandwidth satisfies Eq.(50). This is the premise of constructing an adaptive Cauchy kernel, which will overcome the problem of dynamically distributed noise.

Based on the above results and analysis, the Cauchy kernel bandwidth will be adaptively adjusted via the constraint:

$$\sigma(t) = \varphi(t)\sigma_{\max} \tag{51}$$

$$\varphi(t) = 1 - \exp\left(\frac{\mathbf{P}(t)}{\left(\widehat{\mathbf{y}}(t)\right)^2}\right) \tag{52}$$

where $\varphi(t)$ is the adaptive parameter, $\sigma_{\max}$ is the maximum bandwidth, and is calculated according to Eq.(50).

When estimating the system under the influence of Gaussian noise, $\varphi(t)$ tends to approach 1, which means that the Cauchy kernel will have a wider bandwidth. On the contrary, when there is an impact of an outlier, the innovation value $\widehat{\mathbf{y}}(t)$ will change greatly, which leads to a decrease in the value of $\varphi(t)$. This is the adaptive adjustment mechanism of the Cauchy kernel bandwidth applied in this paper.

### *C. Derivation of the proposed algorithm*

In this section, the AMEEF-DCKF algorithm is developed based on the LFAC algorithm and utilizes an adaptive Cauchy kernel, which overcomes the problem of communication burden and the sensitivity of kernel bandwidth. First, another representation of $\widehat{\mathbf{K}}(t)$ in Eq.(42) after applying the result in Appendix A, is given by:

$$\widehat{\mathbf{K}}(t)=\left(\widehat{\mathbf{P}}(t)+\mathbf{S}^{T}(t)\widehat{\mathbf{P}}_{xy}(t)\right)^{-1}\left(\widehat{\mathbf{P}}_{yx}(t)\widehat{\mathbf{R}}_{yy}^{-1}(t)+\widehat{\mathbf{S}}^{T}(t)\right)\times \left[\begin{array}{l}\widehat{\mathbf{R}}_{yy}^{-1}(t)+\mathbf{S}(t)\left(\widehat{\mathbf{P}}_{xx}(t)+\mathbf{S}^{T}(t)\widehat{\mathbf{P}}_{xy}(t)\right)^{-1}\times \\ \times\left(\widehat{\mathbf{P}}_{yx}(t)\widehat{\mathbf{R}}_{yy}^{-1}(t)+\mathbf{S}^{T}(t)\right)\end{array}\right]^{-1} \tag{53}$$

Since $\mathbf{\Pi}(t)$ is a strictly diagonal matrix, $\mathbf{\Pi}_{yx}(t)$ and $\mathbf{\Pi}_{xy}(t)$ have little impact on the performance of the estimation algorithm. Therefore, some close approximations, including $\mathbf{P}_{yx}(k)=\mathbf{0}_{nxm}$ and $\mathbf{P}_{xy}(k)=\mathbf{0}_{mxn}$, can be applied. At the same time, the process noise $\mathbf{q}(t)$ is considered as a Gaussian distribution, so $\mathbf{P}_{xx}(t)$ can be described by:

$$\widehat{\mathbf{P}}(t)=\left(\mathbf{\Xi}_{p}^{-1}(t|t-1)\right)^{T}\mathbf{\Xi}_{p}(t|t-1)=\widehat{\mathbf{P}}_{xx}^{-1}(t|t-1) \tag{54}$$

The gain matrix $\mathbf{K}(t)$ in Eq.(53) can be rewritten by:

$$\mathbf{K}(t)=\widehat{\mathbf{P}}(t\mid t-1)\mathbf{S}^{T}(t)\left[\mathbf{R}_{yy}^{-1}(t)+\mathbf{S}(t)\widehat{\mathbf{P}}(t\mid t-1)\mathbf{S}^{T}(t)\right]^{-1} \tag{55}$$

Then $\hat{\mathbf{x}}(t|t)$ and $\widehat{\mathbf{P}}(t|t)$ are written as information forms, in which the derivation is based on [18] (Appendix B):

$$\hat{\mathbf{x}}(t\mid t)\approx\widehat{\mathbf{P}}(t\mid t)\left[\widehat{\mathbf{P}}^{-1}(t\mid t-1)\hat{\mathbf{x}}(t\mid t-1)+\mathbf{S}^{T}(t)\mathbf{R}_{yy}(t)\mathbf{y}(t)\right] \tag{56}$$

$$\widehat{\mathbf{P}}(t\mid t)\approx\left[\widehat{\mathbf{P}}^{-1}(t\mid t-1)+\mathbf{S}^{T}(t)\mathbf{R}_{yy}(t)\mathbf{S}(t)\right]^{-1} \tag{57}$$

In a centralized sensor network with $n$ sensors, the information obtained is in the form of: $\mathbf{y}(t)=\left[\mathbf{y}_1^T(t),\mathbf{y}_2^T(t),..,\mathbf{y}_n^T(t)\right]^T$, $\mathbf{S}(t)=\left[\mathbf{S}_1^T(t),\mathbf{S}_2^T(t),..,\mathbf{S}_n^T(t)\right]^T$, and $\mathbf{R}_{yy}(t)=\text{blkdiag}\left(\mathbf{R}_{1,y}(t),\mathbf{R}_{2,y}(t),\ldots,\mathbf{R}_{n,y}(t)\right)$, in which $\mathbf{y}_i(t)$, $\mathbf{S}_i(t)$, and $\mathbf{R}_{i,y}(t)$ are obtained by each sensor itself. Then Eq.(56) and Eq.(57) are rewritten as:

$$\hat{\mathbf{x}}(t\mid t)\approx\widehat{\mathbf{P}}(t\mid t)\left[\widehat{\mathbf{P}}^{-1}(t\mid t-1)\hat{\mathbf{x}}(t\mid t-1)+\sum_{i=1}^{n}\mathbf{S}_i^T\mathbf{R}_{i,yy}(t)\mathbf{y}_i(t)\right] \tag{58}$$

$$\widehat{\mathbf{P}}(t\mid t)\approx\left[\widehat{\mathbf{P}}^{-1}(t\mid t-1)+\sum_{i=1}^{n}\mathbf{S}_i^T(t)\mathbf{R}_{i,yy}(t)\mathbf{S}_i(t)\right]^{-1} \tag{59}$$

It is easy to observe that the direct implementation of Eq. (58) and Eq. (59) in a multi-sensor network would impose a significant communication burden. In order to overcome this problem, the distributed consensus average algorithm LFAC, which we have developed, is applied to achieve consensus among sensors. Based on the developed LFAC algorithm, the representations are obtained as:

$$\begin{cases}\mathbf{D}(t)=\sum_{i=1}^{n}\mathbf{D}_i(t)\\ \mathbf{V}(t)=\sum_{i=1}^{n}\mathbf{V}_i(t)\end{cases} \tag{60}$$

with $\mathbf{D}_i(t)=\mathbf{S}_i^T(t)\mathbf{R}_{i,yy}(t)\mathbf{y}_i(t)$; $\mathbf{V}_i(t)=\mathbf{S}_i^T(t)\mathbf{R}_{i,yy}(t)\mathbf{S}_i(t)$.

Based on the convergence result of Eq.(11), the distributed estimates of $\hat{\mathbf{x}}(t\mid t)$ and $\widehat{\mathbf{P}}(t\mid t)$ are rewritten by:

$$\hat{\mathbf{x}}(t\mid t)\approx\mathbf{P}(t\mid t)\left[\mathbf{P}^{-1}(t\mid t-1)\hat{\mathbf{x}}(t\mid t-1)+\mathbf{D}(t)\right] \tag{61}$$

$$\mathbf{P}(t\mid t)\approx\left[\widehat{\mathbf{P}}^{-1}(t\mid t-1)+\mathbf{V}(t)\right]^{-1} \tag{62}$$

In summary of the above analysis, the pseudocode of the AMEEF-DCKF is shown in **Algorithm 1**.

---

**Algorithm 1** Pseudocode of the AMEEF-DCKF

---

***Step 1:*** Consider a multi-sensor network ($n$ sensor) consisting of $m$ stationary leaders and $n$-$m$ followers.
Set initial: $\hat{\mathbf{x}}_{0|0}$; $\mathbf{P}_{0|0}$; $\delta$ (threshold)
$A$, $A_1$, $A_2$ satisfy Eq.(8); $s(0)=\beta(0)$; $\omega(0)=1_n$; $\gamma=10^{-6}$
***For:*** $t=1,2,3,\ldots$
***Step 2:*** Calculate $\hat{\mathbf{x}}_{t|t-1}$; $\mathbf{P}_{t|t-1}$ through Eq.(21-22)
Calculate $\mathbf{y}(t)$; $\mathbf{H}(t)$ through Eq. (28-29)
***Step 3:*** Performing the Cholesky decomposition $\mathbf{\Xi}(t)$ yields $\mathbf{\Xi}_p(t|t-1)$ and $\mathbf{\Xi}_r(t)$
Calculate $\mathbf{d}(t)$; $\mathbf{W}(t)$ through Eq.(33-34)
***Step 4:*** Set $k=1$; $\hat{\mathbf{x}}^0(t|t)=\hat{\mathbf{x}}(t|t-1)$
Update the Cauchy kernel bandwidth $\sigma_i(t)$ based on Eq.(51)
Calculate $e_i(t)=d_i(t)-\mathbf{w}_i(t)\mathbf{x}(t)$
Calculate $\mathbf{K}(t)$ through Eq.(55)
Calculate $\mathbf{D}(t)$ and $\mathbf{V}(t)$ through Eq.(60)
Calculate $\hat{\mathbf{x}}(t\mid t)$ through Eq.(61)
***If*** $\left\|\hat{\mathbf{x}}^k(t|t)-\hat{\mathbf{x}}^{k-1}(t|t)\right\|/\left\|\hat{\mathbf{x}}^{k-1}(t|t)\right\|\le\delta$ hold,
set $\hat{\mathbf{x}}(t)=\hat{\mathbf{x}}^k(t|t)$ and go to ***step 5***.
***else***, set $k=k+1$, and go back calculate $e_i(t)$
***Step 5:*** Update $\mathbf{P}(t|t)$ through Eq.(62)
***End for***

---

## IV. Performance Analysis

### *A. Convergence proof of the LFAC algorithm*

Let $u_r$ and $u_l$ be two eigenvectors corresponding to the right and left of $A_1$, in which $A_1$ has an eigenvalue equal to 1, obtaining the limit [39]:

$$\lim_{t\to\infty}A_1^t=G \tag{63}$$

where $G=u_r u_l^T$ is the projection matrix onto the eigenspace corresponding to the eigenvalue 1 of $A_1$. In other words, $G$ is the projection onto the null space of $(I_{n-m}-A_1)$. Therefore, Eq.(63) can be rewritten as:

$$\lim_{t\to\infty}\frac{A_1^{t-1}+\ldots+A_1+I_{n-m}}{t}=u_r u_l^T \tag{64}$$

Then, the state update step in the algorithm becomes:

$$\beta_i(t)=\frac{\left[A_1^t\beta_F(0)+\left\{A_1^{t-1}+\ldots+A_1+I_{n-m}\right\}A_2\beta_L(0)\right]_i}{\left[A_1^t 1_{n-m}+\left\{A_1^{t-1}+\ldots+A_1+I_{n-m}\right\}A_2 1_m\right]_i} \tag{65}$$

Divide both the numerator and denominator by $t$, and take the limit $\beta_i^s=\lim_{t\to\infty}\beta_i(t)$, to get:

$$\beta_i^s = \frac{\left[\lim_{t\to\infty}\frac{A_1^t\beta_F(0)}{t} + \lim_{t\to\infty}\left\{\frac{A_1^{t-1}+...+A_1+I_{n-m}}{t}\right\}A_2\beta_L(0)\right]_i}{\left[\lim_{t\to\infty}\frac{A_1^t 1_{n-m}}{t} + \lim_{t\to\infty}\left\{\frac{A_1^{t-1}+...+A_1+I_{n-m}}{t}\right\}A_2 1_m\right]_i} \tag{66}$$

Realizing that $\left[\lim_{t\to\infty}\frac{A_1^t\beta_F(0)}{t}\right]_i \to 0$, and $\left[\lim_{t\to\infty}\frac{A_1^t 1_{n-m}}{t}\right]_i \to 0$, thus:

$$\beta_i^s = \frac{\left[\lim_{t\to\infty} A_1^t A_2 \beta_L(0)\right]_i}{\left[\lim_{t\to\infty} A_1^t A_2 1_m\right]_i} = \frac{\left[u_r 1_{n-m}^T A_2 \beta_L(0)\right]_i}{\left[u_r 1_{n-m}^T A_2 1_m\right]_i} \tag{67}$$

On the other hand, since $A_2$ is column stochastic, i.e. $1_{n-m}^T A_2 = 1_m^T$. Therefore:

$$\beta_i^s = \frac{\left[u_r 1_m^T \beta_L(0)\right]_i}{\left[u_r 1_m^T 1_m\right]_i} = \frac{\left[u_r 1_m^T \beta_L(0)\right]_i}{\left[u_r m\right]_i} = \frac{1_m^T \beta_L(0)}{m} =: \bar{\beta}_L(0) \tag{68}$$

Thus, the followers' state values converge to the average of the leaders' state values.

### B. Computational Complexity Analysis

The computational complexity at each step of the proposed algorithm AMEEF-DCKF is shown in TABLE I. It is noted that $m$ and $n$ represent the dimensions of the measurement vector $\mathbf{y}(t)$ and the state vector $\mathbf{x}(t)$, respectively. The highest order of magnitude with $m < n$ is calculated and considered as the total cost. $T_e$ denotes the computational burden on a scalar number of the exponential function.

TABLE I

COMPUTATIONAL COMPLEXITY OF AMEEF-DCKF

| Computation | Overall cost |
|---|---|
| $\mathbf{S}(t)$ in Eq.(25) | $O(n^3)$ |
| $\mathbf{\Xi}_p(t\|t-1)$ in Eq.(27) | $O(2n^3)$ |
| $\mathbf{\Xi}_r(t)$ in Eq.(27) | $O(2m^3)$ |
| $e_i(t)$ in Eq.(28) | $O(2n^3)$ |
| $\mathbf{\Pi}(t)$ in Eq.(36) | $O(2n^2T_e)$ |
| $\hat{\mathbf{x}}(t\|t)$ in Eq.(57) | $O(2nm)$ |
| $\mathbf{P}(t\|t)$ in Eq.(58) | $O(4n^3)$ |
| $\mathbf{K}(t)$ in Eq.(51) | $O(4n^3)$ |

Based on the results given in TABLE I, it can be observed that the computational complexity of every variable in the proposed algorithm does not exceed $O(4n^3)$, i.e., the computational cost of AMEEF-DCKF is three orders of magnitude per iteration. Thus, this result is consistent with that of the conventional CKF. The additional computational burden of the gain matrix $\mathbf{K}(t)$ with weight coefficients including $\mathbf{\Xi}_p(t|t-1)$, $\mathbf{\Xi}_r(t)$, $\mathbf{\Pi}(t)$ and $\mathbf{S}(t)$, is the difference between the proposed algorithm and the traditional CKF. Since in the traditional CKF, the covariance matrix $\mathbf{P}(t)$ and the Kalman gain $\mathbf{K}(t)$ are also calculated, the proposed algorithm AMEEF-DCKF has a higher computational complexity by an additional amount $T_{MEEF-DACKF}^{add}$ compared to the CKF:

$$\begin{aligned} T_{MEEF-DACKF}^{add} &= T_{MEEF-DACKF}^{\sum} - T_{CKF}^{\sum} \\ &= \theta\left(2n^2T_e + 2m^3 + 8n^3\right) \\ &\approx \theta\left(2n^2T_e + 8n^3\right) \quad (m \ll n) \end{aligned} \tag{69}$$

where $\theta$ is the average fixed-point iteration number; $T_{MEEF-DACKF}^{\sum}$ : total computational burden of AMEEF-DCKF; $T_{CKF}^{\sum}$ : total computational burden of CKF.

## V. Simulation Results

In this section, experiments are performed to estimate the state of two typical nonlinear objects: power systems and land vehicles, to demonstrate the excellent performance of the proposed algorithm in complex environments. The performance was evaluated using a Monte Carlo method with 200 independent runs, each containing 500 samples. The root mean square error (RMSE) evaluation criteria used:

$$\mathrm{RMSE}(t) = \sqrt{\frac{1}{200}\sum_{i=1}^{200}\left\|\hat{\mathbf{x}}_i(t) - \mathbf{x}_i(t)\right\|_2^2} \tag{70}$$

where $\mathbf{x}(t)$ and $\hat{\mathbf{x}}(t)$ are true value and estimated value of the state, respectively. The average RMSE (ARMSE) is defined as $\mathrm{ARMSE} = \frac{1}{500}\sum_{t=1}^{500}\mathrm{RMSE}(t)$.

In this paper, several noise models, including Gaussian, mixed-Gaussian, bimodal Gaussian mixture with outliers, and Rayleigh noise, are employed to evaluate their impact on the accuracy of state estimates across two applications: power systems (voltage magnitude and angle) and land vehicle navigation (position and velocity).

① The Gaussian noise [14, 15, 17]:

$$r \sim G(a, \ell) \tag{71}$$

where $G(a,\ell)$ is the Gaussian distribution with variance $\ell$ and mean $a$; this type of Gaussian noise is represented as $r_G \sim G(a,\ell)$.

② The mixed-Gaussian noise [4, 15, 17, 25, 33, 40]:

$$r \sim \tau G(a,\ell_1) + (1-\tau)G(a,\ell_2) \tag{72}$$

where $\tau$ is the mixture coefficient; this type of Gaussian noise is represented as $r_{mG} \sim mG(\tau, a, \ell_1, \ell_2)$.

③ The Rayleigh noise [4, 40]:

$$r(t) = \frac{t}{\varphi^2}\exp\left(-\frac{t^2}{2\varphi^2}\right) \tag{73}$$

where $\varphi$ is the shape parameter. This type of Rayleigh noise is represented as $r_{Rayleigh} \sim Ray(\varphi)$.

④ Bimodal Gaussian mixture noise and outliers [14, 22]:

$$r \sim \rho G(a,\ell_1)+\nu G(s,\ell_2)+\rho G(a,\ell_3) \;\; ;(2\rho+\nu=1) \qquad (74)$$

where $G(s,\ell_2)$ is the Gaussian distribution with variance $\ell_2$ and mean $s$; this type of bimodal Gaussian mixture noise and outliers is represented as $r_{bmG} \sim bmG(\rho,\nu,a,s,\ell_1,\ell_2,\ell_3)$.

In addition, the communication topology of the multi-sensor network [9,10] used in this paper is depicted in Figure 1, which distinguishes the leader and the follower.

The simulation program is run on a Core™ i7-5600U-CPU 2.60GHz computer. In addition, DCKF [13], MEE-DCKF [4], MCC-DCKF [18], and MEEF-DCKF [22] distributed algorithms were implemented to compare their performance with the proposed algorithms. It should be noted that all these algorithms use Cauchy kernels, and the values of the coefficients are provided in TABLE II. Regarding the issue of setting the values of these coefficients, they are directly referenced in studies [4, 14, 15, 17, 22, 25].

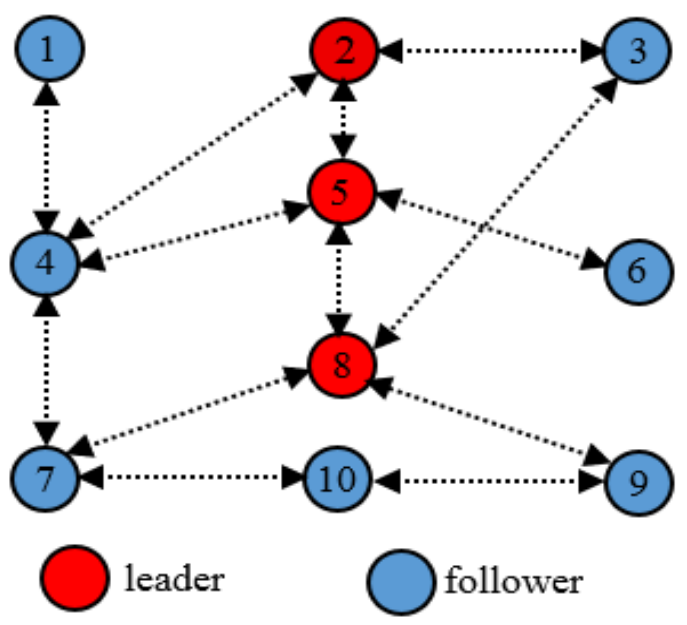

Fig. 1. Communication topology of leader-follower sensors.

TABLE II
COEFFICIENTS OF THE DISTRIBUTED ALGORITHMS

| Coefficient / Algorithm | Mixture coefficient | Threshold | Kernel width |
|---|---|---|---|
| MCC-DCKF | ./. | $\delta = 10^{-6}$ | σ=25; |
| MEE-DCKF | ./. | $\delta = 10^{-6}$ | σ=6; |
| MEEF-DCKF | $\eta = 0.5$ | $\delta = 10^{-6}$ | $\sigma_1$=1.8; $\sigma_2$=1.8; |
| AMEEF-DCKF | $\eta = 0.5$ | $\delta = 10^{-6}$ | ./. |

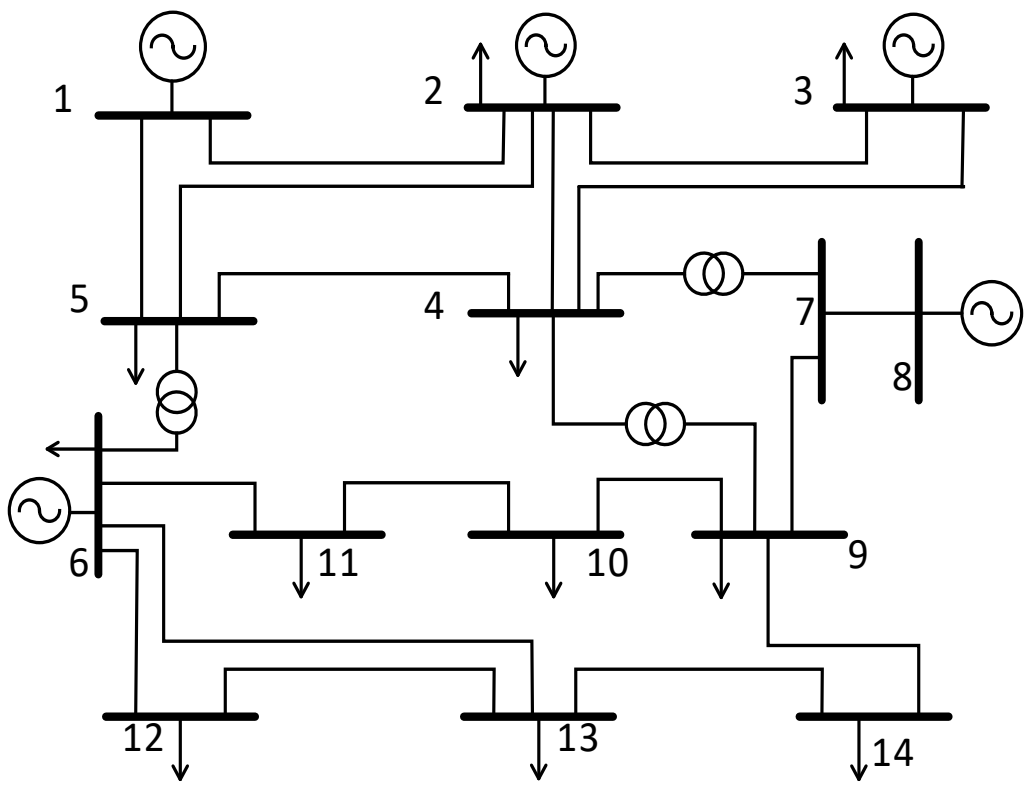

Fig. 2. Structure of IEEE 14-bus system.

### *A. Power system state estimation*

For the task of estimating the state of the power system, the performance of AMEEF-DCKF is verified on an IEEE 14-bus system, where the structure of the IEEE 14-bus system, illustrated in Figure 2 consists of three transformers, five wind generators, 20 branches, and 14 buses.

The state equation of the system is given by:

$$\mathbf{x}(t)=\mathbf{x}(t-1)+\mathbf{q}(t-1) \qquad (75)$$

where $\mathbf{x}(t)=\left[\{Y_f\}_{i=1}^{14},\{\varphi_f\}_{i=1}^{14}\right]$ with $Y_f$, $\varphi_f$ are the magnitude and phase of voltage at bus $f$, respectively; $\mathbf{q}(t)$ is the Gaussian noise with zero mean, covariance matrix $\mathbf{Q}(t)=10^{-5}\mathbf{I}_n$.

The measurements include: reactive power injection $P_f$, real power injection $Q_f$, voltage amplitude $Y_f$, reactive power flow $P_{f\text{-}j}$, and real power flow $Q_{f\text{-}j}$, which are given by [33]:

$$\begin{cases} Q_f = \sum_{j=1}^{N} Y_f Y_j \left(B_{fj}\cos\varphi_{fj} + U_{fj}\sin\varphi_{fj}\right) \\ P_f = \sum_{j=1}^{N} Y_f Y_j \left(B_{fj}\sin\varphi_{fj} - U_{fj}\cos\varphi_{fj}\right) \\ Q_{f-j} = Y_f^2\left(B_{gf}+B_{fj}\right) - Y_f Y_j\left(B_{fj}\cos\varphi_{fj} + U_{fj}\sin\varphi_{fj}\right) \\ P_{f-j} = -Y_f^2\left(U_{gf}+U_{fj}\right) - Y_f Y_j\left(B_{fj}\sin\varphi_{fj} - U_{fj}\cos\varphi_{fj}\right) \end{cases} \qquad (76)$$

where $\varphi_{fj}$ is the voltage angle between bus $j$ and $f$. $U_{fj}$, $U_{gj}$, $B_{fj}$, and $B_{gj}$ are the susceptances and conductances, respectively.

Under the influence of noise given by Eq.(72), the actual measurement obtained per node via a 10-node sensor network is as follows:

$$\begin{cases} \mathbf{y}_i(t) = \mathbf{Z}_1 + \mathbf{r}_i(t) & ;i=1,3,5,7,9 \\ \mathbf{y}_i(t) = \mathbf{Z}_2 + \mathbf{r}_i(t) & ;i=2,4,6,8,10 \end{cases} \qquad (77)$$

where $\mathbf{Z}_1$, $\mathbf{Z}_2$ are described in TABLE III; $\mathbf{r}_i(t)$ is the Gaussian noise with zero mean, covariance matrix $\mathbf{R}_i(t)=10^{-4}\mathbf{I}_m$.

TABLE III
MEASUREMENT INFORMATION IN $\mathbf{Z}_1$ AND $\mathbf{Z}_2$

| | |
|---|---|
| $\mathbf{Z}_1$ | $Y_1$, $\varphi_1$, $Q_2$, $Q_4$, $Q_6$, $Q_8$, $Q_{10}$, $Q_{12}$, $Q_{14}$, $P_2$, $P_4$, $P_6$, $P_8$, $P_{10}$, $P_{12}$, $P_{14}$, $Q_{1\text{-}5}$, $Q_{2\text{-}4}$, $Q_{3\text{-}4}$, $Q_{4\text{-}7}$, $Q_{5\text{-}6}$, $Q_{6\text{-}12}$, $Q_{7\text{-}8}$, $Q_{9\text{-}10}$, $Q_{10\text{-}11}$, $Q_{13\text{-}14}$, $P_{1\text{-}5}$, $P_{2\text{-}4}$, $P_{3\text{-}4}$, $P_{4\text{-}7}$, $P_{5\text{-}6}$, $P_{6\text{-}12}$, $P_{7\text{-}8}$, $P_{9\text{-}10}$, $P_{10\text{-}11}$, $P_{13\text{-}14}$ |
| $\mathbf{Z}_2$ | $Y_1$, $\varphi_1$, $Q_4$, $Q_5$, $Q_6$, $Q_8$, $Q_{10}$, $Q_{11}$, $Q_{12}$, $Q_{14}$, $P_4$, $P_5$, $P_6$, $P_8$, $P_{10}$, $P_{11}$, $P_{12}$, $P_{14}$, $Q_{1\text{-}2}$, $Q_{1\text{-}5}$, $Q_{2\text{-}3}$, $Q_{2\text{-}4}$, $Q_{2\text{-}5}$, $Q_{4\text{-}7}$, $Q_{6\text{-}11}$, $Q_{6\text{-}13}$, $Q_{12\text{-}13}$, $P_{1\text{-}2}$, $P_{1\text{-}2}$, $P_{1\text{-}5}$, $P_{2\text{-}3}$, $P_{2\text{-}4}$, $P_{2\text{-}5}$, $P_{4\text{-}7}$, $P_{6\text{-}11}$, $P_{6\text{-}13}$, $P_{12\text{-}13}$ |

The initial sate $\mathbf{x}(0)$ is provided by [41], with $\mathbf{x}(0|0)=[1_{1\times14},0_{1\times14}]^T$ and $\mathbf{P}(0|0)=\mathbf{I}_n$.

Next, one by one, the scenarios impacting the IEEE 14-bus system are tested to demonstrate the performance of the proposed algorithm.

***Scenario 1***: $r_G \sim G(0,10^{-2})$ noise

In this scenario, both the measurement and process noise are assumed to be Gaussian. The results of estimating the amplitude and angle state of the voltage at bus-7 and node-5 of the algorithms are illustrated in Figures 3 and 4. Based on the estimated results obtained, it can be observed that the performance of the algorithms is comparable. This result accurately reflects the assumptions used in this paper, which is that in the face of Gaussian distributed noise, the bandwidth size of the adaptive Cauchy kernel will be set at a high value.

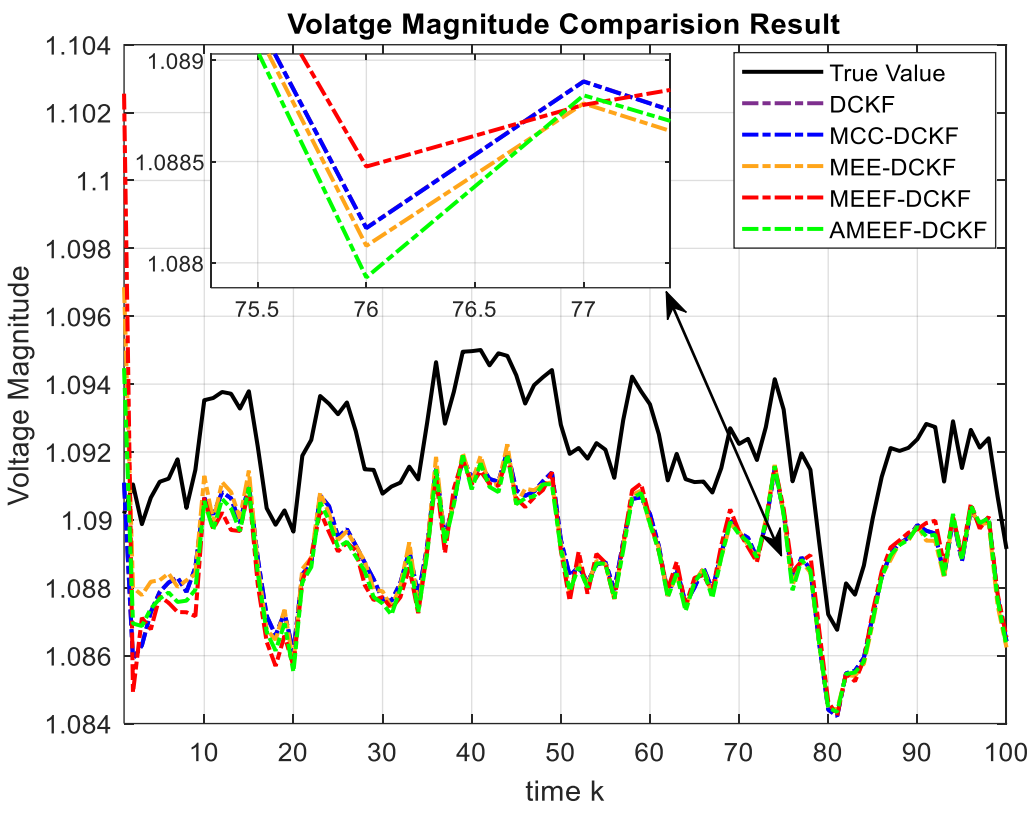

Fig. 3. Voltage magnitude at bus-7 and node-5 for scenario 1

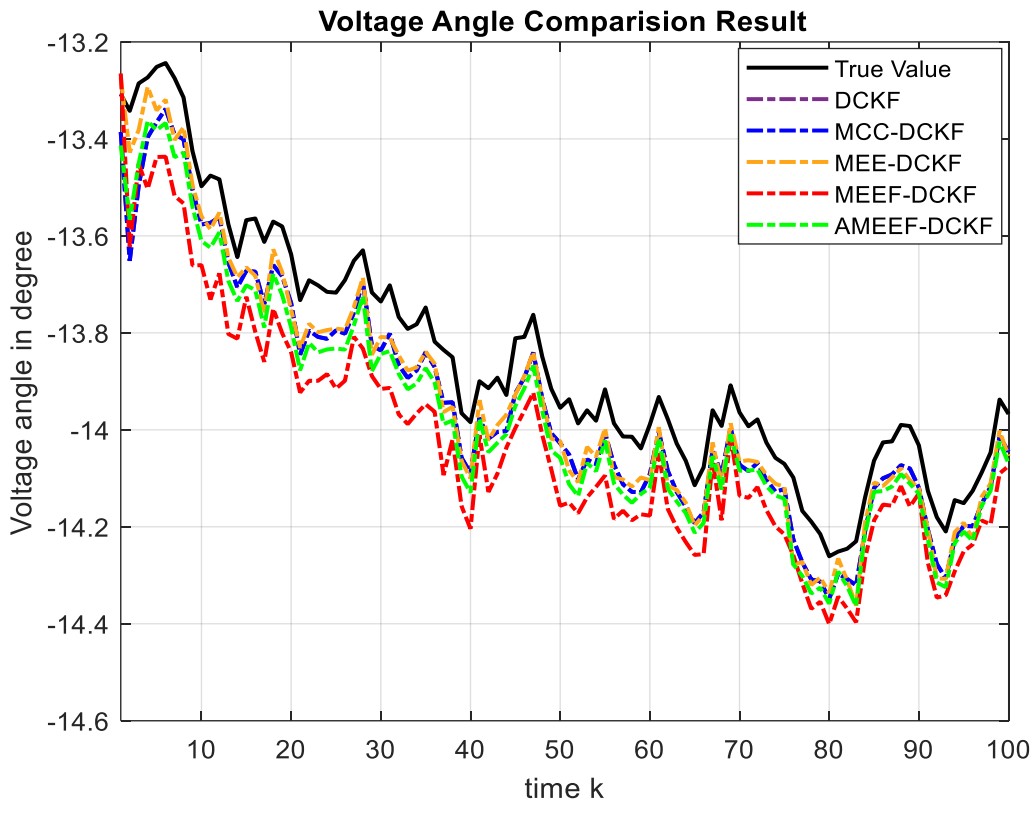

Fig. 4. Voltage angle at bus-7 and node-5 for scenario 1

***Scenario* 2**: Abnormal data and $r_{Rayleigh} \sim Ray(3)$ noise

In this scenario, the power injection assumption of bus-7 and bus-8 with a deviation of ΔP7=-1 and ΔP8=-1 is established, in which the acquired data of sensor node-5 is used. Besides, the power system is also affected by $r_{Rayleigh}$ noise. The results of the estimation of the amplitude and angle states of the voltage at bus-7 are illustrated in Figures 5 and 6, respectively. At the same time, ARMSE results are provided in TABLE IV.

Based on the results obtained, it can be easily observed that the proposed algorithm continues to show its robustness in this complex scenario. Under the influence of abnormal data and Rayleigh noise, the MEE-DCKF algorithm has revealed its poor numerical stability, while MEEF-DCKF, although it has given good estimation results, still does not meet the current requirements. In contrast, the AMEEF-DCKF algorithm still shows the best performance thanks to the flexibility of the adaptive Cauchy kernel. Specifically, the estimation accuracy of voltage magnitude is 14%, 87%, 86% and 91% higher than MEEF-DCKF, MEE-DCKF, MCC-DCKF, and DCKF, respectively.

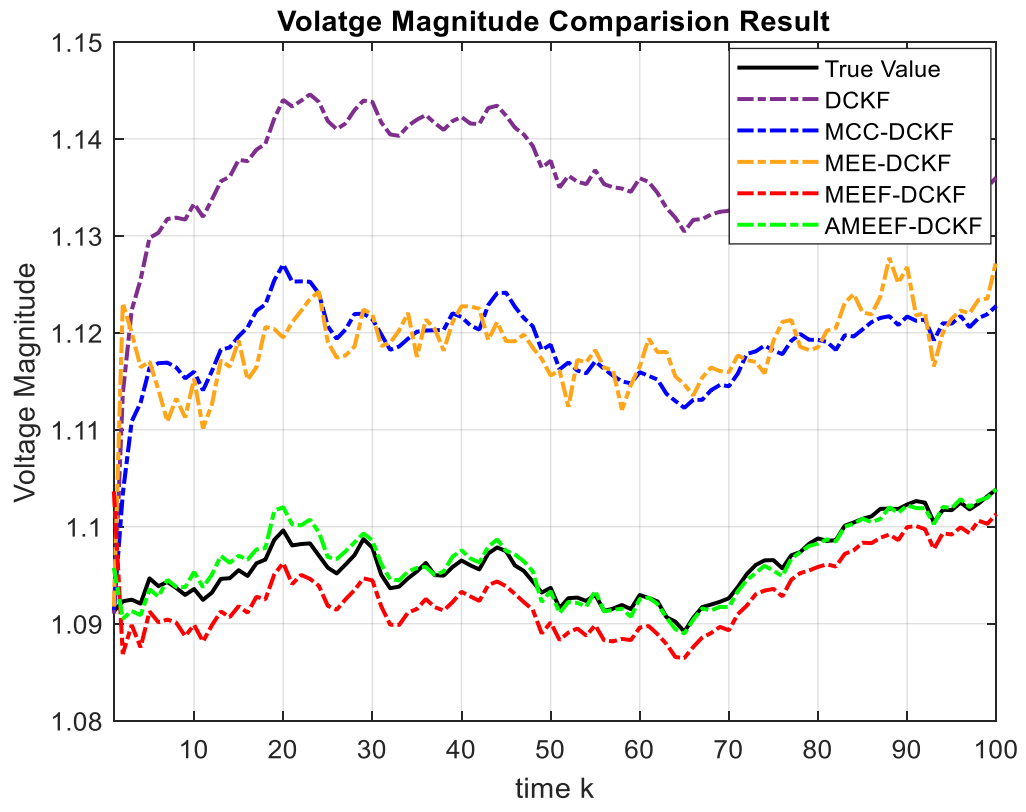

Fig. 5. Voltage magnitude at bus-7 and node-5 for scenario 2

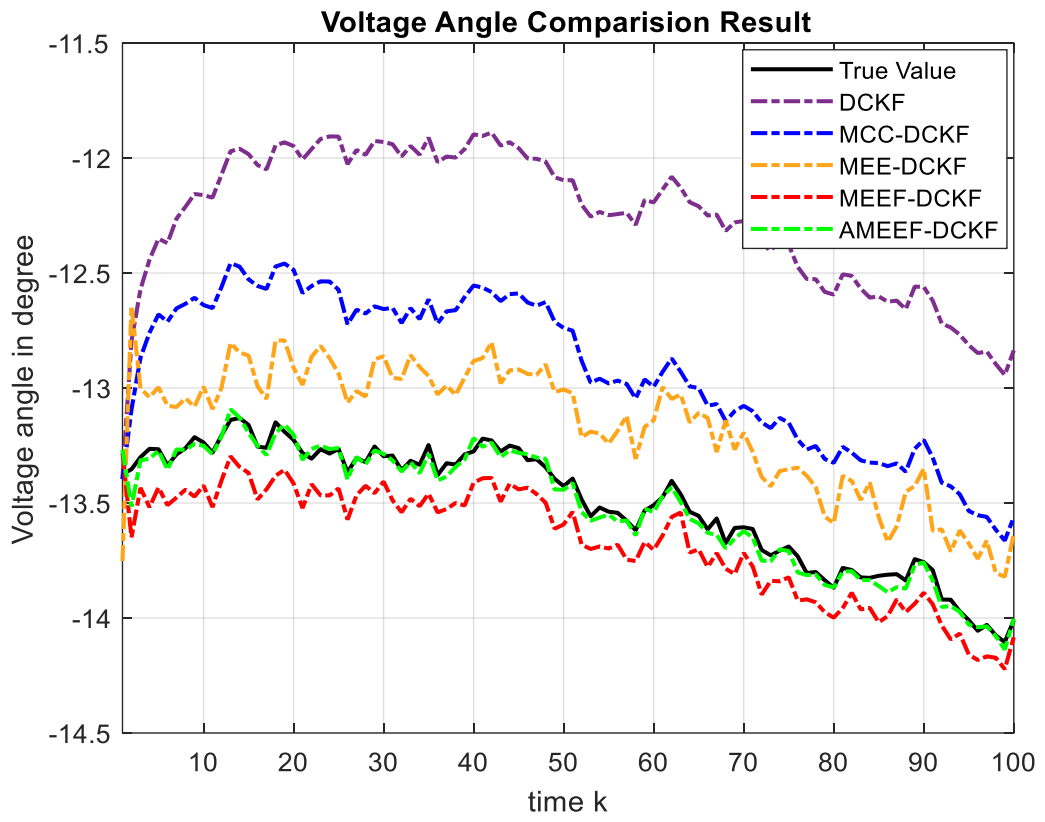

Fig. 6. Voltage angle at bus-7 and node-5 for scenario 2

TABLE IV

ARMSE OF VOLTAGE ANGLE AND MAGNITUDE FOR SCENARIO 2

| ARMSE / Algorithm | Voltage magnitude | Voltage angle |
|---|---|---|
| DCKF | 0.2091 | 0.0909 |
| MCC-DCKF | 0.1345 | 0.0710 |
| MEE-DCKF | 0.1425 | 0.0449 |
| MEEF-DCKF | 0.0220 | 0.0210 |
| AMEEF-DCKF | **0.0190** | **0.0137** |

***Scenario* 3**: Sudden load change and $r_{Rayleigh} \sim Ray(5)$ noise

A complex scenario is investigated, in which at the 30th test time, due to the change in load power, the voltage amplitude suddenly drops by 15%, while being affected by the noise $r_{Rayleigh} \sim Ray(5)$, to verify the robustness and accuracy of the proposed algorithm. The state estimation results are illustrated in Figures 7 and 8.

It can be easily observed that the proposed algorithm achieves superior accuracy and stability compared to DCKF, MCC-DCKF, MEE-DCKF, and MEEF-DCKF, especially at the time of sudden changes. More specifically, in the estimation of voltage phase angle at this time, the graphs depicting the estimation results of DCKF, MCC-DCKF, MEE-DCKF, and MEEF-DCKF algorithms show large bumps. This is a potentially dangerous problem in applications requiring high accuracy.

Additionally, for a more comprehensive comparison, the single-step running time of the estimation algorithms is described in Table V. It can be observed that AMEEF-DCKF requires a larger computation time. However, given the excellent accuracy and robustness it achieves, this trade-off is completely acceptable.

TABLE V

COMPARE SINGLE-STEP RUNNING TIME (s)

| Scenario / Algorithm | Scenario 1 | Scenario 2 | Scenario 3 |
|---|---|---|---|
| DCKF | 0.0076 | 0.0082 | 0.0087 |
| MCC-DCKF | 0.0093 | 0.0098 | 0.0095 |
| MEE-DCKF | 0.0111 | 0.0137 | 0.0141 |
| MEEF-DCKF | 0.0137 | 0.0169 | 0.0171 |
| AMEEF-DCKF | 0.0162 | 0.0186 | 0.0192 |

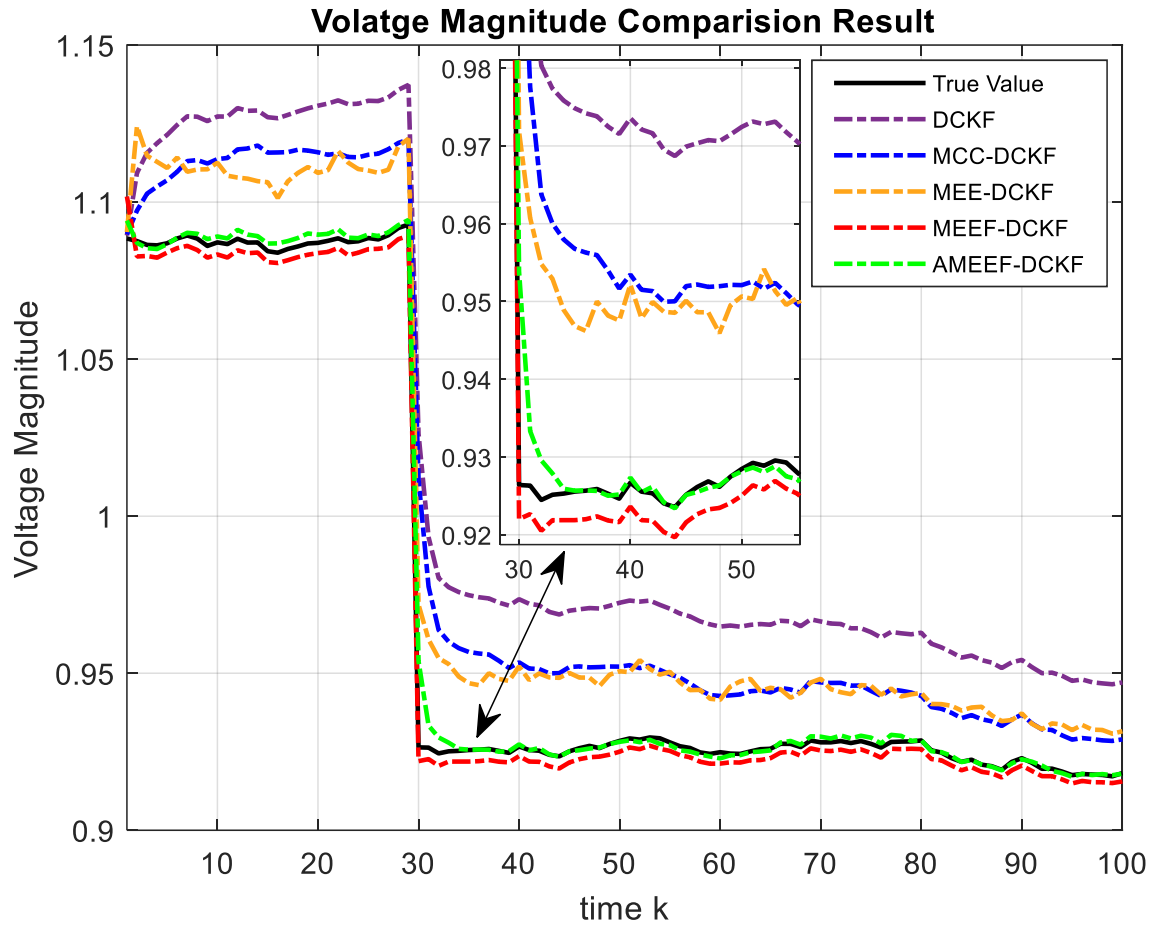

Fig. 7. Voltage magnitude at bus-7 and node-5 for scenario 3

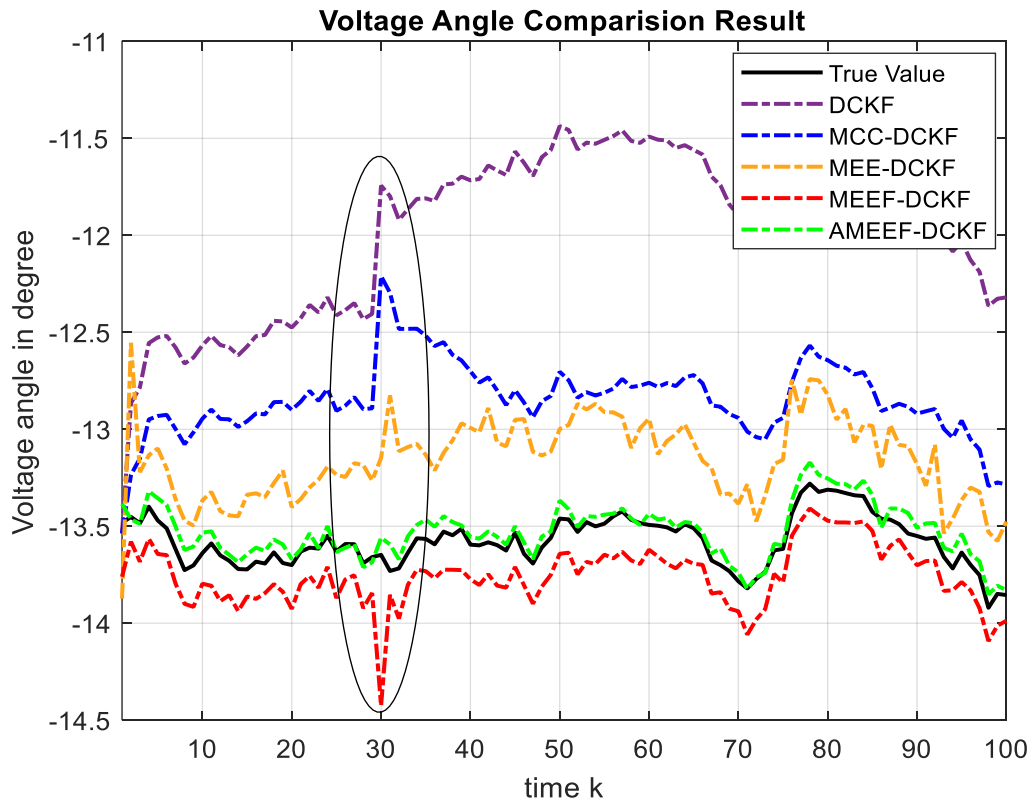

Fig. 8. Voltage angle at bus-7 and node-5 for scenario 3

Considering the results provided in Tables I and V together, it can be seen that although the complexity and computation time per step are higher than the traditional CKF, the proposed algorithm achieves superior nonlinear noise handling and stability, especially in complex, realistic scenarios. With suitable hardware and optimization strategies, such as using modern platforms (GPUs, FPGAs) or adjusting the fixed-point iteration threshold (δ), the proposed algorithm AMEEF-DCKF will fully meet the accuracy and timing requirements in today's real-time applications.

### *B. Land vehicle state estimation*

For land vehicles, the same model as in [17] is used, which is given by:

$$\mathbf{x}(t)=\begin{bmatrix}1&0&\Delta t&0\\0&1&0&\Delta t\\0&0&1&0\\0&0&0&1\end{bmatrix}\mathbf{x}(t-1)+\mathbf{q}(t-1) \quad (78)$$

where $\mathbf{x}(t)=[x_1(t),x_2(t),x_3(t),x_4(t)]^T$ is the state vector comprising of north and east position, the north and east velocity.

The initial state $\mathbf{x}(0)$ is set as $\mathbf{x}(0)=[0,0,5,5]^T$, the sampling interval is set as $\Delta t=0.3$(s). The process noise $\mathbf{q}(t)=[q_1(t),q_2(t),q_3(t),q_4(t)]^T$ is obeying zero-mean Gaussian distribution, $q_1(t)\sim G(0,0.01)$, $q_2(t)\sim G(0,0.01)$, $q_3(t)\sim G(0,1)$, and $q_4(t)\sim G(0,1)$.

The observation models of $n$=10 sensors are:

$$\mathbf{y}_i(t)=\begin{bmatrix}-1&0&-1&0\\0&-1&0&-1\end{bmatrix}\mathbf{x}(t)+\mathbf{r}_i(t) \quad ;\ i=1,3,5,7,9 \quad (79)$$

$$\mathbf{y}_i(t)=\begin{bmatrix}-1&0&0&0\\0&-1&0&0\end{bmatrix}\mathbf{x}(t)+\mathbf{r}_i(t) \quad ;i=2,4,6,8,10 \quad (80)$$

The initial state $\mathbf{x}(0|0)$ of all nodes is set to $\mathbf{x}(0|0)=[1,1,1,1]^T$ with $\mathbf{P}(0|0)=[900,900,4,4]^T$.

Next, similar to the power system example, different noise scenarios are tested one by one to illustrate the performance of the proposed algorithm.

***Scenario 4***: $r_{bmG}\sim bmG\left(0.4,0.2,0.2,0,10^{-2},20,0.3\right)$ noise

In this scenario, the effect of the symmetrically mixed Gaussian noise $r_{bmG}$ (multi-peak distribution) on the land vehicle is considered, in which the Gaussian distribution with small probability and large variance can create impulsive noises [14]. The RMSE results of the position and velocity estimates of node-5 are illustrated in Figures 9 and 10.

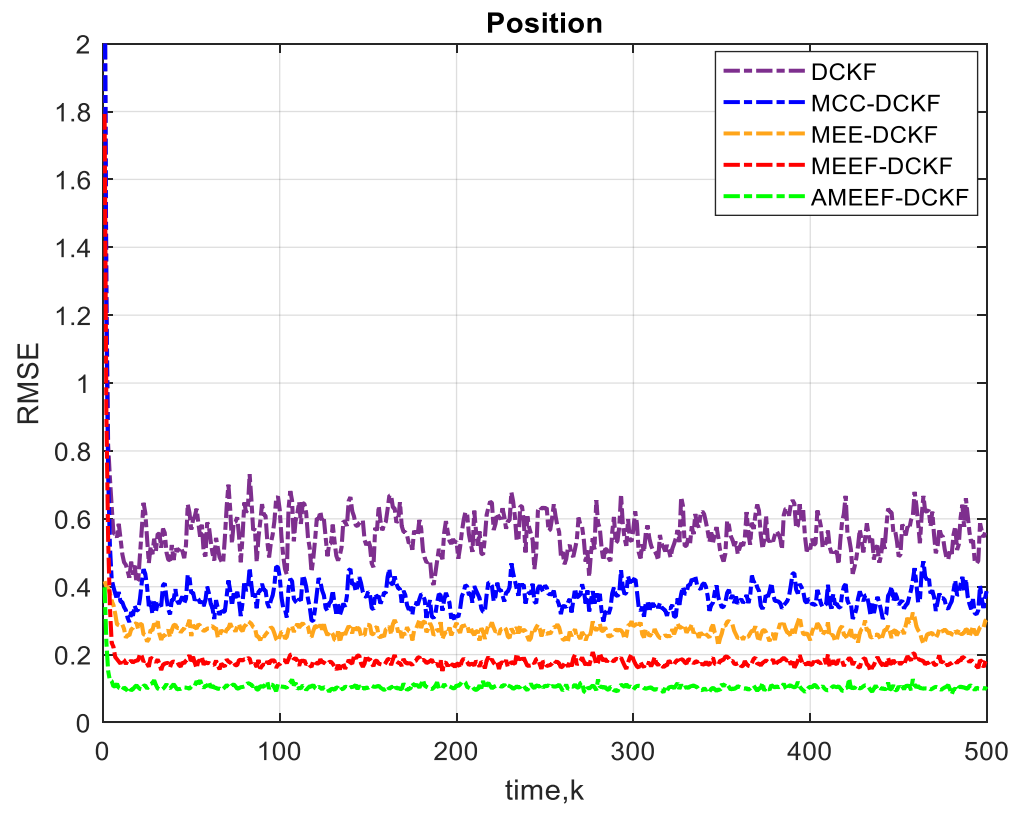

Fig. 9. RMSE of position at node-5 for scenario 4

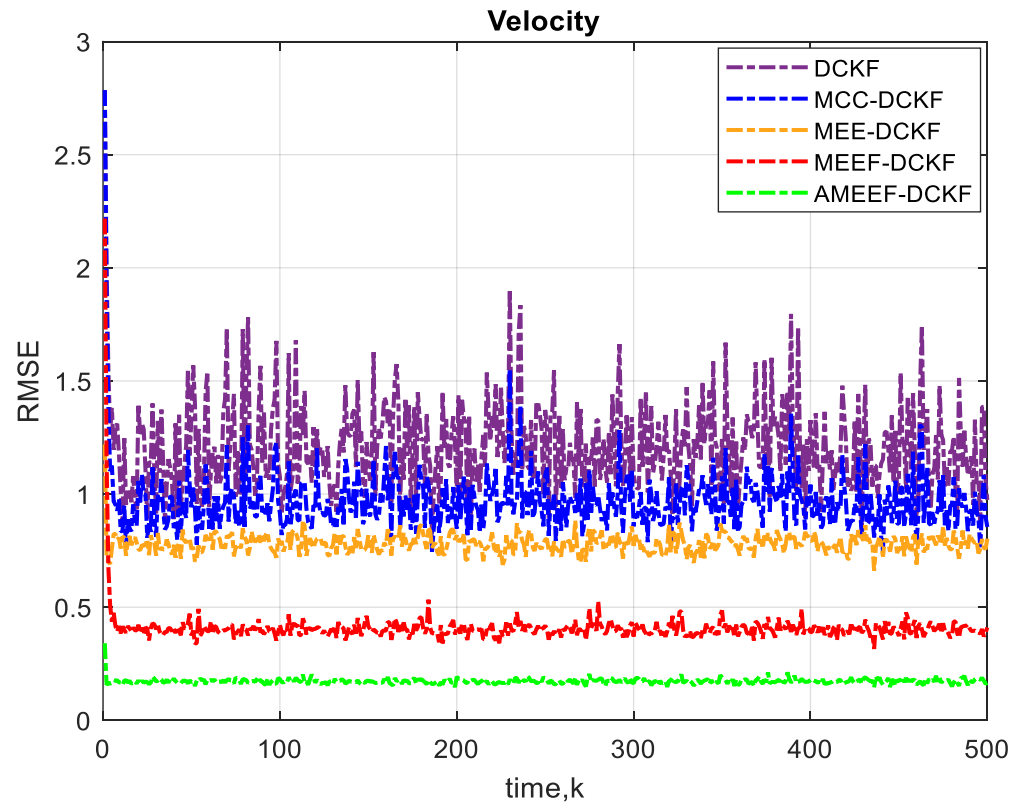

Fig. 10. RMSE of velocity at node-5 for scenario 4

Based on the performance obtained, it can be concluded that thanks to the error PDF's ability to automatically locate the peak at zero, MEEF-DCKF and AMEEF-DCKF achieve better-estimated performance than MEE-DCKF, MCC-DCKF, and DCKF. It is also easy to observe that the proposed algorithm still obtains the best results. Specifically, its position estimation performance is 48%, 66%, 75%, and 83% higher than MEEF-DCKF, MEE-DCKF, MCC-DCKF, and DCKF, respectively.

***Scenario* 5**: $r_{mG} \sim mG\left(0.6, 0.2, 10^{-4}, 10^{-2}\right)$ noise

In this scenario, the effect of $r_{mG}$ noise on the land vehicle is considered. The ARMSE of the position and velocity of node-5 is shown in Figure 11. Because of the flexibility of the adaptive Cauchy kernel, AMEEF-DCKF continues to exhibit the best performance. Specifically, its estimated velocity efficiency is 64%, 82%, 85%, and 89% higher than MEEF-DCKF, MEE-DCKF, MCC-DCKF, and DCKF, respectively.

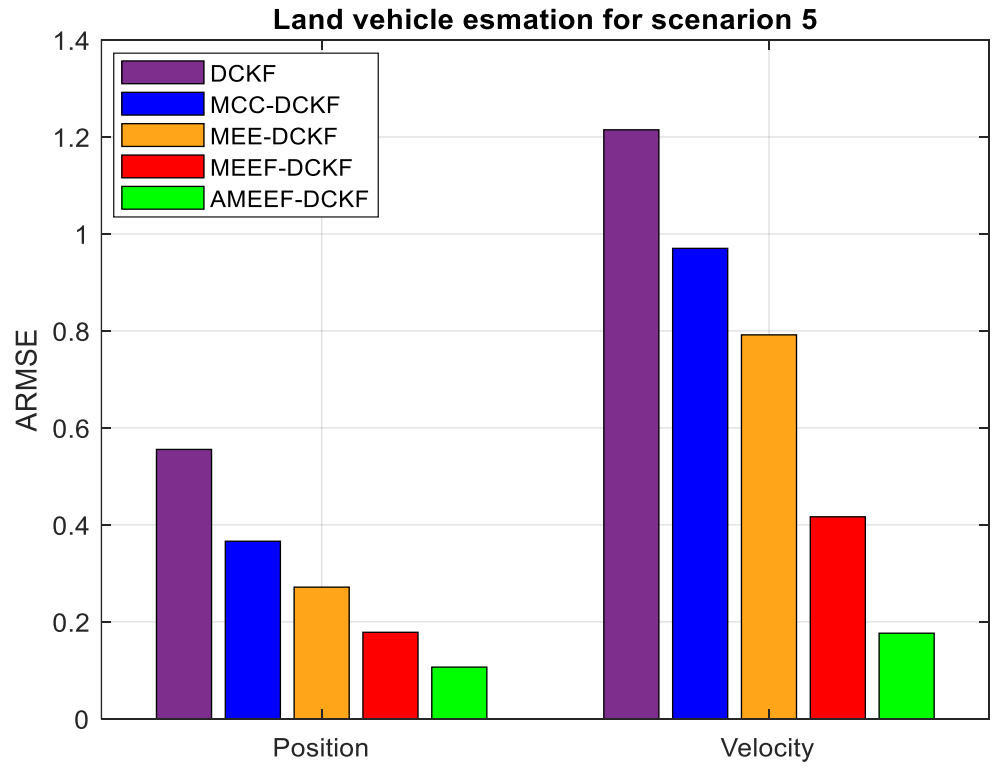

Fig. 11. ARMSE of velocity and position at node-5 for scenario 5

In scenarios 4 and 5, the performance of the proposed algorithm has been validated through two error evaluation criteria, including RMSE and ARMSE. To further emphasize the superior accuracy of the proposed algorithm compared to existing algorithms, the mean absolute error (MAE) ( $\mathrm{MAE}(t) = \frac{1}{200}\sum_{i=1}^{200}\left|\hat{\mathbf{x}}_i(t) - \mathbf{x}_i(t)\right|$ ) evaluation criterion was chosen for additional comparison. Specifically, the MAE of velocity and position is provided in Table VI, which clearly shows that AMEEF-DCKF continues to demonstrate superior accuracy compared to the algorithms being compared. Furthermore, this result confirmed the claim made in recently published studies that MEEF-based filters achieve better performance than MEE, which in turn is better than MCC.

TABLE VI
MAE OF VELOCITY AND POSITION

| Algorithm \ Scenario | Scenario 4 | | Scenario 5 | |
|---|---|---|---|---|
| | position | velocity | position | velocity |
| DCKF | 0.600550 | 1.281389 | 0.730429 | 1.437657 |
| MCC-DCKF | 0.211227 | 0.627930 | 0.371477 | 0.714894 |
| MEE-DCKF | 0.194726 | 0.422003 | 0.205403 | 0.594661 |
| MEEF-DCKF | 0.067184 | 0.110737 | 0.075971 | 0.144512 |
| AMEEF-DCKF | **0.033068** | **0.077679** | **0.041544** | **0.090538** |

## VI. Conclusion

In this paper, stemming from the problems faced when estimating the state and navigation of nonlinear objects including non-Gaussian noise, abnormal data, and communication burden, the AMEEF-DCKF algorithm has been proposed to overcome the above limitations. In which, the LFAC algorithm has been developed to overcome the problem of communication burden in multi-sensor networks. The follower sensors have been reached consensus with the average value of the leader sensor through the Push-Sum protocol. In addition, under the optimal criterion constructed AMEEF, in which the kernels are adaptive Cauchy kernels, the influence of non-Gaussian noise, abnormal data, the singular matrix and the sensitivity of the kernel bandwidth have been processed easily. Based on the simulation results obtained in the scenario of two nonlinear objects, the power system and the land vehicle, the excellent performance of the proposed algorithm has been confirmed.

## Appendix A

Consider Eq.(38), let $\tilde{\mathbf{P}}(t) = \hat{\mathbf{P}}_{xx}(t) + \mathbf{S}^T(t)\hat{\mathbf{P}}_{xy}(t)$ and $\tilde{\mathbf{S}}(t) = \hat{\mathbf{P}}_{yx}(t)\hat{\mathbf{R}}_{yy}^{-1}(t) + \mathbf{S}^T(t)$ , then this equation can be rewritten as:

$$\hat{\mathbf{K}}(t) = \left[\tilde{\mathbf{P}}(t) + \tilde{\mathbf{S}}(t)\hat{\mathbf{R}}_{yy}(t)\mathbf{S}(t)\right]^{-1}\tilde{\mathbf{S}}(t)\hat{\mathbf{R}}_{yy}(t) \tag{81}$$

Applying the matrix inverse lemma yields:

$$(A + BCD)^{-1} = A^{-1} - A^{-1}B(DA^{-1}B + C^{-1})^{-1}DA^{-1} \tag{82}$$

where $A = \tilde{\mathbf{P}}(t)$, $B = \tilde{\mathbf{S}}(t)$, $C = \mathbf{R}_{yy}(t)$, $D = \mathbf{S}(t)$

Then $\tilde{\mathbf{P}}(t) + \tilde{\mathbf{S}}(t)\hat{\mathbf{R}}_{yy}(t)\mathbf{S}(t)$ can be simplified as

$$\begin{aligned}&\tilde{\mathbf{P}}(t) + \tilde{\mathbf{S}}(t)\hat{\mathbf{R}}_{yy}(t)\mathbf{S}(t) = \\ &= \tilde{\mathbf{P}}^{-1}(t) - \tilde{\mathbf{P}}^{-1}(t)\tilde{\mathbf{S}}(t)\left[\mathbf{S}(t)\tilde{\mathbf{P}}^{-1}(t)\tilde{\mathbf{S}}(t) + \hat{\mathbf{R}}_{yy}^{-1}(t)\right]^{-1}\mathbf{S}(t)\tilde{\mathbf{P}}^{-1}(t)\end{aligned} \tag{83}$$

Then Eq.(81) can be calculated by:

$$
\begin{aligned}
&\widehat{\mathbf{K}}(t)=\tilde{\mathbf{P}}^{-1}(t)\tilde{\mathbf{S}}(t)\widehat{\mathbf{R}}_{yy}(t)-\tilde{\mathbf{P}}^{-1}(t)\tilde{\mathbf{S}}(t)\times\\
&\quad\times\left[\mathbf{S}(t)\tilde{\mathbf{P}}^{-1}(t)\tilde{\mathbf{S}}(t)+\widehat{\mathbf{R}}_{yy}^{-1}(t)\right]^{-1}\mathbf{S}(t)\tilde{\mathbf{P}}^{-1}(t)\tilde{\mathbf{S}}(t)\widehat{\mathbf{R}}_{yy}(t)\\
&=\tilde{\mathbf{P}}^{-1}(t)\tilde{\mathbf{S}}(t)\left[\mathbf{S}(t)\tilde{\mathbf{P}}^{-1}(t)\tilde{\mathbf{S}}(t)+\widehat{\mathbf{R}}_{yy}^{-1}(t)\right]^{-1}\times\\
&\quad\times\left[\left(\mathbf{S}(t)\tilde{\mathbf{P}}^{-1}(t)\tilde{\mathbf{S}}(t)+\widehat{\mathbf{R}}_{yy}^{-1}(t)\right)\widehat{\mathbf{R}}_{yy}(t)-\mathbf{S}(t)\tilde{\mathbf{P}}^{-1}(t)\tilde{\mathbf{S}}(t)\widehat{\mathbf{R}}_{yy}(t)\right]\\
&=\tilde{\mathbf{P}}^{-1}(t)\tilde{\mathbf{S}}(t)\left[\mathbf{S}(t)\tilde{\mathbf{P}}^{-1}(t)\tilde{\mathbf{S}}(t)+\widehat{\mathbf{R}}_{yy}^{-1}(t)\right]^{-1}
\end{aligned}
\tag{84}
$$

Substituting $\tilde{\mathbf{P}}(t)$ and $\tilde{\mathbf{S}}(t)$ into Eq.(84), Eq.(53) is obtained.

## APPENDIX B

The Eq.(44) can be rewritten as:

$$
\begin{aligned}
&\widehat{\mathbf{P}}(t|t)=\widehat{\mathbf{P}}(t|t-1)-\widehat{\mathbf{P}}(t|t-1)\mathbf{S}^T(t)\widehat{\mathbf{K}}^T(t)-\\
&-\widehat{\mathbf{K}}(t)\mathbf{S}(t)\widehat{\mathbf{P}}(t|t-1)+\widehat{\mathbf{K}}(t)\left[\mathbf{S}(t)\widehat{\mathbf{P}}(t|t-1)\mathbf{S}^T(t)+\mathbf{R}(t)\right]\widehat{\mathbf{K}}^T(t)
\end{aligned}
\tag{85}
$$

According to Eq.(55), Eq.(85) can be approximated as:

$$
\begin{aligned}
\widehat{\mathbf{P}}(t|t)&\approx\widehat{\mathbf{P}}(t|t-1)-\widehat{\mathbf{P}}(t|t-1)\mathbf{S}^T(t)\widehat{\mathbf{K}}^T(t)\\
&\quad-\widehat{\mathbf{K}}(t)\mathbf{S}(t)\widehat{\mathbf{P}}(t|t-1)+\widehat{\mathbf{P}}(t|t-1)\mathbf{S}^T(t)\widehat{\mathbf{K}}^T(t)\\
&=\widehat{\mathbf{P}}(t|t-1)-\widehat{\mathbf{K}}(t)\mathbf{S}(t)\widehat{\mathbf{P}}(t|t-1)\\
&=\widehat{\mathbf{P}}(t|t-1)-\widehat{\mathbf{P}}(t|t-1)\mathbf{S}^T(t)\left[\mathbf{R}^{-1}(t)+\mathbf{S}(t)\widehat{\mathbf{P}}(t|t-1)\mathbf{S}^T(t)\right]^{-1}\\
&\quad\times\mathbf{S}(t)\widehat{\mathbf{P}}(t|t-1)
\end{aligned}
\tag{86}
$$

The matrix inverse lemma applied, Eq.(86) can be expressed as

$$
\widehat{\mathbf{P}}(t\,|\,t)=\left(\widehat{\mathbf{P}}^{-1}(t\,|\,t-1)+\widehat{\mathbf{S}}^T(t)\mathbf{R}_{yy}(t)\widehat{\mathbf{S}}(t)\right)^{-1}\tag{87}
$$

In addition, the gain matrix $\widehat{\mathbf{K}}(t)$ can be expressed as:

$$
\widehat{\mathbf{K}}(t)=\widehat{\mathbf{P}}(t\,|\,t)\widehat{\mathbf{S}}^T(t)\mathbf{R}_{yy}(t)\tag{88}
$$

Then $\hat{\mathbf{x}}(t|t)$ can be calculated as:

$$
\begin{aligned}
&\hat{\mathbf{x}}(t\,|\,t)=\hat{\mathbf{x}}(t\,|\,t-1)+\widehat{\mathbf{P}}(t\,|\,t)\mathbf{S}^T(t)\mathbf{R}_{yy}(t)\left(\mathbf{y}(t)-\mathbf{S}(t)\hat{\mathbf{x}}(t\,|\,t-1)\right)\\
&=\widehat{\mathbf{P}}(t\,|\,t)\left[\widehat{\mathbf{P}}^{-1}(t\,|\,t)\hat{\mathbf{x}}(t\,|\,t-1)+\mathbf{S}^T(t)\mathbf{R}_{yy}(t)\left(\mathbf{y}(t)-\mathbf{S}(t)\hat{\mathbf{x}}(t\,|\,t-1)\right)\right]\\
&=\widehat{\mathbf{P}}(t\,|\,t)\begin{bmatrix}\left(\widehat{\mathbf{P}}^{-1}(t\,|\,t-1)+\mathbf{S}^T(t)\mathbf{R}_{yy}(t)\mathbf{S}(t)\right)\hat{\mathbf{x}}(t\,|\,t-1)\\+\mathbf{S}^T(t)\mathbf{R}_{yy}(t)\left(\mathbf{y}(t)-\mathbf{S}(t)\hat{\mathbf{x}}(t\,|\,t-1)\right)\end{bmatrix}\\
&=\widehat{\mathbf{P}}(t\,|\,t)\left[\widehat{\mathbf{P}}^{-1}(t\,|\,t-1)\hat{\mathbf{x}}(t\,|\,t-1)+\mathbf{S}^T\mathbf{R}_{yy}(t)\mathbf{y}(t)\right]
\end{aligned}
\tag{89}
$$

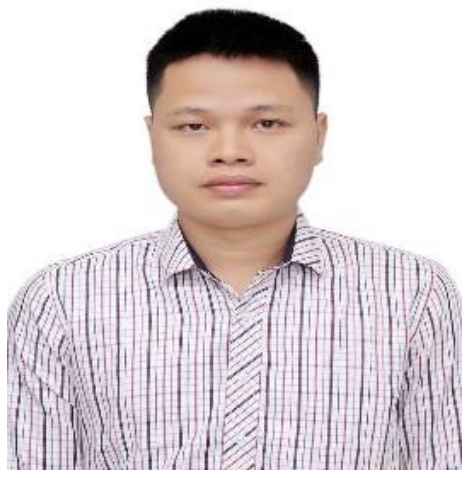

**Duc Viet Nguyen** received a master's degree in control engineering and automation from Le Quy Don University, Viet Nam, in 2019. He is working toward the Ph.D. degree in signal and information processing from the School of Electrical Engineering at Southwest Jiaotong University, Chengdu, China.

His current research interests include designing automatic control systems, state estimation, and adaptive filtering algorithms.

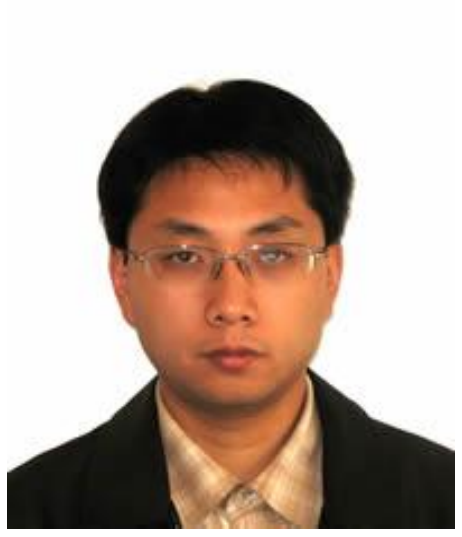

**Haiquan Zhao** (Senior Member, IEEE) received the B.S. degree in applied mathematics and the M.S. and Ph.D. degrees in signal and information processing from Southwest Jiaotong University, Chengdu, China, in 1998, 2005, and 2011, respectively.

Since 2012, he has been a Professor with the School of Electrical Engineering, Southwest Jiaotong University. From 2015 to 2016, he was a Visiting Scholar with the University of Florida, Gainesville, FL, USA. He is the author or coauthor of more than 280 international journal papers (SCI indexed) and owns 56 invention patents. His current research interests include information theoretical learning, adaptive filters, adaptive networks, active noise control, Kalman filters, machine learning, and artificial intelligence.

Dr. Zhao has won several provincial and ministerial awards and many best paper awards at international conferences or IEEE TRANSACTIONS. He has served as an Active Reviewer for several IEEE TRANSACTIONS, IET series, signal processing, and other international journals. He is currently a Handling Editor of Signal Processing, and also an Associate Editor for IEEE Transaction on Audio, Speech and Language Processing, IEEE TRANSACTIONS ON SYSTEMS, MAN AND CYBERNETICS: SYSTEM, IEEE SIGNAL PROCESSING LETTERS, IEEE SENSORS JOURNAL, and IEEE OPEN JOURNAL OF SIGNAL PROCESSING.

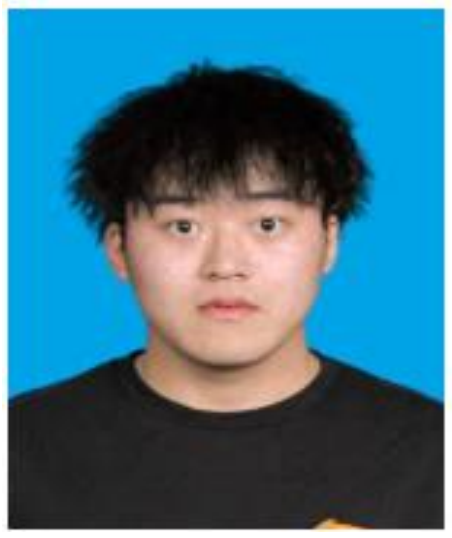

**Jinhui Hu** received the B.E. degree in electrical engineering and automation from Chang' An University, Xi'an, China, in 2022. He is currently working toward the Ph.D degree in signal and information processing from the School of Electrical Engineering, Southwest Jiaotong University, Chengdu, China.

His current research interests include state estimation and adaptive filtering algorithms.